\documentclass[]{rsproca_new}

\usepackage{amsmath}
\usepackage{amssymb}
\usepackage{bm}
\usepackage{graphicx}
\usepackage{subcaption}

\usepackage[numbers,sort&compress]{natbib}

\begin{document}

\title{Theory of the flow-induced deformation of shallow compliant microchannels with thick walls}

\author{Xiaojia Wang and Ivan C.\ Christov}

\address{School of Mechanical Engineering, Purdue University, West Lafayette, Indiana 47907, USA}

\subject{mechanical engineering, microsystems, applied mathematics}


\corres{Ivan C.\ Christov\\
\email{christov@purdue.edu}}

\begin{abstract}
Long, shallow microchannels embedded in thick soft materials are widely used in microfluidic devices for lab-on-a-chip applications. However, the bulging effect caused by fluid--structure interactions between the internal viscous flow and the soft walls has not been completely understood. Previous models either contain a fitting parameter or are specialized to channels with plate-like walls. This work is a theoretical study of the steady-state response of a compliant microchannel with a thick wall. Using lubrication theory for low-Reynolds-number flows and the theory for linearly elastic isotropic solids, we obtain perturbative solutions for the flow and deformation. Specifically, only the channel's top wall deformation is considered, and {the ratio between its thickness $t$ and width $w$ is assumed to be $(t/w)^2 \gg 1$}. We show that the deformation at each stream-wise cross-section can be considered independently, and that the top wall can be regarded as a simply supported rectangle subject to uniform pressure at its bottom. The stress and displacement fields are found using  Fourier series, {based on which} the channel shape and the hydrodynamic resistance are calculated, yielding a new flow rate--pressure drop relation without fitting parameters. Our results agree favorably with, and thus rationalize, previous experiments.
\end{abstract}


\begin{fmtext}

\section{Introduction}
\label{sec:intro}
Microfluidic devices have enabled the miniaturization of  processes that involve the flow and manipulation of small volumes of fluids, down to the nanoliter \cite{SQ05}. In recent years, polymer-based materials have become popular for the fabrication of microfluidic devices because they promise cheaper and faster production cycles \cite{B08}. A widely-used material for microfabrication is
\end{fmtext}
\maketitle
\noindent
the elastomer poly(dimethylsiloxane) (PDMS) \cite{XW98,MW02,SMMC11}. The development of PDMS-based microfluidic devices is intimately related to the emergence of the cutting-edge field known as \emph{lab-on-a-chip} \cite{SSA04,C13_book}. Microchannels made from PDMS have found many applications, such as platforms for organ-on-a-chip models \cite{Huhetal10} and various biological studies (e.g., assays and combinatorial screening) \cite{SW03}. PDMS, also known commercially as Sylgard 184, typically has a low tensile modulus \cite{JMTT14}, and PDMS-based micro-conduits are prone to deformation (``bulging'') under applied forces. Although early studies considered deformation to be a drawback, because it may restrict the structural viability of a device \cite{DSMB97,HJLK02}, the compliance of PDMS microchannels has been exploited to design microfluidic devices with specific functions, such as pressure-actuated valves \cite{TMQ02}, passive fuses \cite{GMV17}, pressure sensors \cite{HHM02,OYE13}, strain sensors \cite{Dhong18}, and impedance-based flow meters with improved sensitivity \cite{NNBR17}.

The soft nature of microfluidic devices made from PDMS or similar materials requires the consideration of \emph{fluid--structure interactions} (FSIs), even at the low Reynolds numbers encountered at these scales \cite{DS16}. Given the broad range of applications of these microsystems, understanding the fluid--structure interplay is important and necessary \cite{KCC18}. Conceptually speaking, the soft wall of the microchannel will bulge due to the hydrodynamic pressure within, and this effect will, in turn, modify the flow velocity and pressure gradient. Rectangular microchannels embedded in soft materials have been the most common configuration. However, even within this ``simple'' Cartesian geometry, quantifying the FSI loop is not easy.  To this end, in the present work, {we analyze the steady-state FSI} in a long and shallow compliant microchannel with thick walls. Gervais \textit{et al.}~\cite{GEGJ06} were the first to put forward a model for this case by introducing a fitting parameter to be determined via calibration with experiments. In their model, the strain in the elastic solid is taken to be linearly proportional to the hydrodynamic pressure, with the proportionality constant being unknown. The same scaling analysis was generalized to the unsteady case by Dendukuri \textit{et al.}~\cite{DGPHD07} for studying stop-flow lithography in a thick-walled microfluidic device. Mukherjee \textit{et al.}~\cite{MCC13} then considered actuation of the soft wall via electroosmotic flow, eliminating the fitting parameter under the assumption of two-dimensional (2D) kinematics (i.e., no displacement in the spanwise direction transverse to the flow). Without the assumption of 2D kinematics, the cross-sectional deformation profile in the spanwise direction was previously assumed to be parabolic. In this way, these studies were able to quantify the flow rate--pressure drop relation and show that this relation deviates from the linear proportionality predicted by Poiseuille’s law. Although the model in \cite{GEGJ06} was initially developed for microchannels with walls behaving like a semi-infinite medium, the model has actually (with varying degrees of success) been also applied to FSI in microchannels with elastic walls of various thicknesses \cite{HUZK09,CTS12,RDC17}.

The fitting parameter in the model from  \cite{GEGJ06} is inconvenient in applications because it has to be recalibrated for each microchannel's geometric and material properties. This model also neglects many fluid--solid coupling details because a solution of the elasticity problem has not been obtained from the governing equations. It has been previously noticed that the fitting parameter has to be related to the top wall thickness \cite{HUZK09}. Seker \textit{et al.}~\cite{SLHLUB09} argued that half-space-like thick walls and plate-like thin walls should be treated differently, and they used a empirical formula to determine the fitting parameter in these two regimes. Later, Christov \textit{et al.}~\cite{CCSS17} used perturbation methods and successfully obtained a fitting-parameter-free model for the flow rate--pressure drop relation in a microchannel with a clamped plate-like top wall. This work was then extended by Shidhore and Christov \cite{SC18} to account for the deformation of microchannels whose top wall behaves like a thick plate. Most recently, Mart\'{i}nez-Calvo \textit{et al.}~\cite{MCSPS19} considered the unsteady version of the problem in \cite{CCSS17} in order to update the previous fitting-parameter-based approach \cite{DGPHD07}. This line of research has shown that fitting parameters are not necessary to quantify microscale FSIs, and they can be avoided entirely by seeking a perturbative solution to the related elasticity problem. All of these models mentioned are applicable to Newtonian fluid flow only. Other works have analyzed FSIs in microchannels conveying non-Newtonian fluids \cite{ADC18,RCDC18}.

All of the existing fitting-parameter-free models target microchannels with \emph{thin} walls (i.e., plate-like elastic structures) and, thus, fail immediately when pushed to a large-thickness regime. Yet, standard fabrication techniques based on soft lithography \cite{XW98,SMMC11} naturally produce microchannels with thick elastic slabs as the boundary, violating both the plate-like assumption and the clamped-boundary assumption. The present work seeks to fill this knowledge gap between theory and experiment. To this end, we derive a new mathematical model (without fitting parameters) for the steady fluid--structure interaction between a Newtonian fluid and an (initially) rectangular microchannel with deformable walls. Specifically, via a scaling analysis of the equilibrium stress balance in the elastic solid, we give a precise definition of ``thick-walled'' microchannels in an asymptotic sense (section \ref{sec:solid_scaling}). The reduced solid mechanics problem is then solved exactly by Fourier series (section \ref{sec:solid_solution}), yielding the precise shape of the fluid--solid interface profile (section \ref{sec:interface}). Next, the flow rate--pressure drop relation is calculated (section \ref{sec:resistance}) by substituting the deformation profile into the leading-order flow field under the lubrication approximation (section \ref{sec:fluids}). The final mathematical result for the nonlinear hydrodynamic resistance is compared with experimental measurements reported in \cite{GEGJ06, RDC17}, and it is found to be in favorable agreement (section \ref{sec:results}). Finally, conclusions and avenues for future work are discussed in section \ref{sec:conclusion}. For completeness, Appendix is provided discussing how the thick-slab elasticity problem limits onto the plate-like problem from previous theoretical studies.

\subsection*{Problem statement, geometry and notation}
\label{sec:problem}
\begin{figure}[ht]
    \centering
    \includegraphics[width=0.75\textwidth]{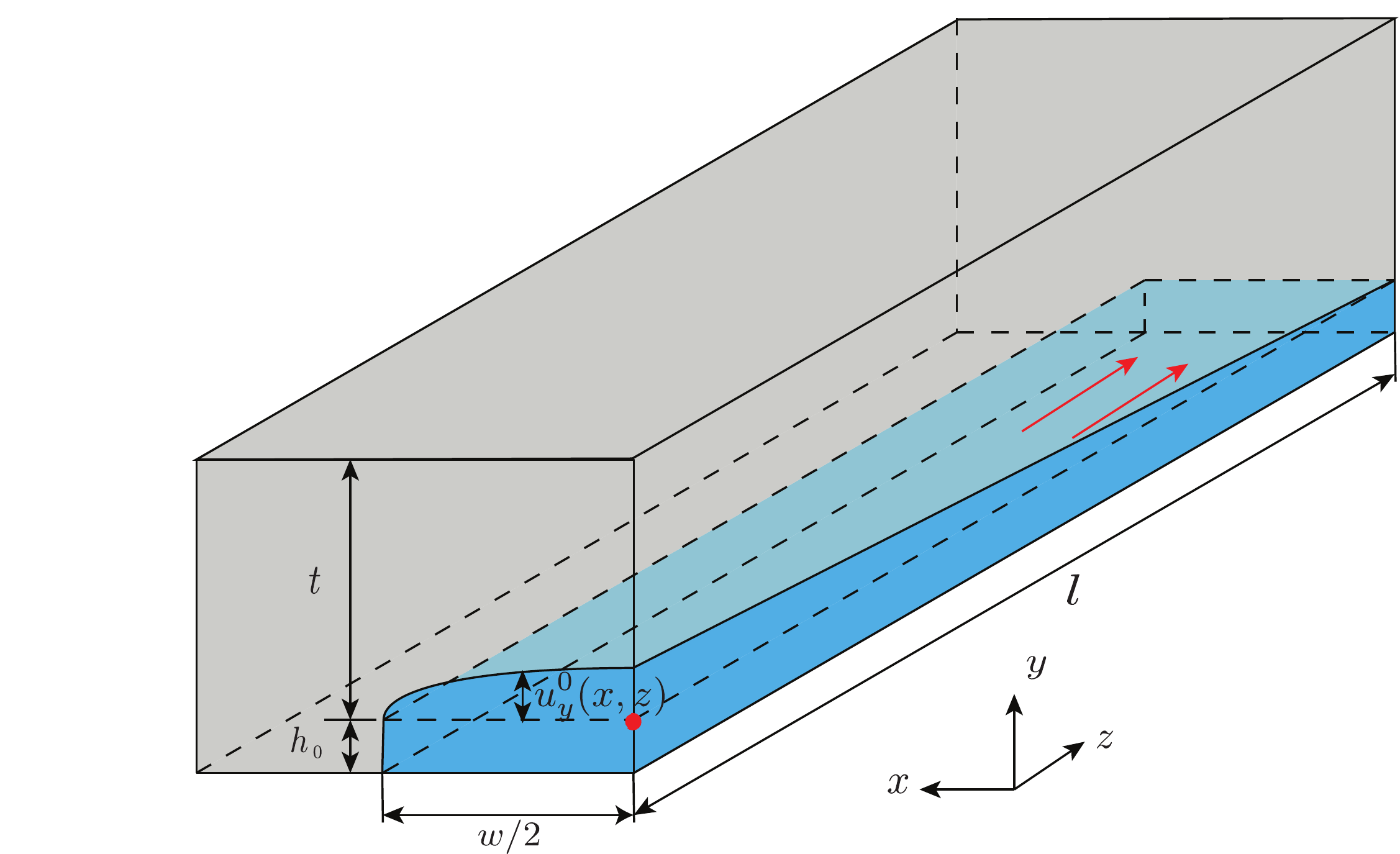}
    \caption{Schematic diagram of the compliant microchannel with key variables labeled. {The origin of the coordinate system (labeled with a red a dot) is set at the centerline ($x=0$) of the undeformed fluid--solid interface, which is initially a distance $h_0$ above the rigid bottom channel wall}. The microchannel is symmetric about $x=0$ thus, for clarity, only half the channel (for $x\geq 0$) is shown. The deformed fluid--solid interface is denoted by the compliant top wall's $y$-displacement evaluated at $y=0$, i.e., $u_y^0(x,z)$. The fluid flow is in the $+z$-direction, as indicated by arrows, with an inlet at $z=0$ and an outlet at $x=l$.}
    \label{fig:schematic}
\end{figure}

Consider the pressure-driven, steady flow of a Newtonian viscous fluid inside of a microchannel fabricated via soft lithographic techniques from an elastomer. A schematic configuration is shown in figure~\ref{fig:schematic}. The problem is symmetric about $x=0$, with only half of it being shown in the figure. The fluid domain within microchannel has undeformed height $h_0$, width $w$ and length $l$. The flow is in the positive $z$-direction. It is assumed that the microchannel is long and shallow, such that $h_0 \ll w \ll l$. Three sides (two lateral walls and the top wall) of the channel are composed of a soft elastic material, while the bottom wall is assumed to be rigid. Here, $t$  denotes the thickness of the top solid slab, while the thickness of the lateral walls is assumed to be large enough to be considered as infinite. According to the scaling analysis of Gervais \textit{et al.}~\cite{GEGJ06}, {the strain of the displaced fluid--solid interface is proportional to the imposed stress, so $u_y^0(x,z) \propto wp(z)/E_Y$, where $p(z)$ denotes the local pressure at flow-wise position $z$, and $E_Y$ denotes the Young's modulus of the solid. Similarly, the side wall deformation is proportional to $h_0 p(z)/E_Y$.} In the regime $h_0\ll w$, the deformation of the top wall is thus expected to be much larger than that of the lateral walls. Therefore, we only consider the deformation of the top wall.

\section{Governing equations of the fluid mechanics problem}
\label{sec:fluids}
First, let us introduce the dimensionless variables
\begin{equation}\label{nd_vars}
    X = \frac{x}{w},\quad Y = \frac{y}{h_0},\quad Z = \frac{z}{l},\quad
    V_X =\frac{\delta v_x}{\epsilon\mathcal{V}_c},\quad  V_Y =\frac{v_y}{\epsilon\mathcal{V}_c},\quad  V_Z =\frac{v_z}{\mathcal{V}_c},\quad P = \frac{p}{\mathcal{P}_0},
\end{equation}
where $\epsilon\mathcal{V}_c/\delta$, $\epsilon\mathcal{V}_c$ and $\mathcal{V}_c$ are respectively the characteristic scales for the velocity components $v_x$, $v_y$ and $v_z$, chosen to balance the conservation of mass equation~\eqref{COM}. Meanwhile, $\mathcal{P}_0$ is the characteristic pressure scale. If the pressure is controlled at both the inlet and outlet, then $\mathcal{P}_0=\Delta p=p(z=l)-p(z=0)$, which is the total pressure drop over the channel length $l$, and $\mathcal{V}_c=h_0^2\Delta p/(\mu l)$. Instead, if {the volumetric flow rate $q$} is controlled at the inlet (and the flow is steady, as assumed), then $\mathcal{V}_c$ is chosen as the mean flow velocity, $\langle v_z\rangle=q/(h_0w)$, and $\mathcal{P}_0=\mu \langle v_z \rangle l/h_0^2$; see the detailed discussion in \cite{CCSS17}. We have also defined the dimensionless ratios $\epsilon = h_0/l$, $\delta = {h_0}/w$, with $\epsilon \ll \delta \ll 1$ (recall section \ref{sec:intro}), and the Reynolds number $Re=\rho\mathcal{V}_c h_0/{\mu}$, where $\rho$ and $\mu$ are the fluid density and dynamic viscosity, respectively. Assuming the working fluid is incompressible, Newtonian, and steady, then the governing equations are the steady incompressible Navier--Stokes equations, which take the dimensionless form:
\begin{subequations}\label{iNS}
\begin{align}
    \frac{\partial V_X}{\partial X}+\frac{\partial V_Y}{\partial Y}+\frac{\partial V_Z}{\partial Z}&=0,\label{COM}\displaybreak[3]\\
    \frac{\epsilon^3}{\delta^2}Re\left(V_X\frac{\partial V_X}{\partial X}+V_Y\frac{\partial V_X}{\partial Y}+V_Z\frac{\partial V_X}{\partial Z}\right)&=-\frac{\partial P}{\partial X}+\epsilon^2\frac{\partial^2 V_X}{\partial X^2}+\frac{\epsilon^2}{\delta^2}\frac{\partial^2 V_X}{\partial Y^2}+\frac{\epsilon^4}{\delta^2}\frac{\partial^2 V_X}{\partial Z^2},\\
    {\epsilon^3}Re\left(V_X\frac{\partial V_Y}{\partial X}+V_Y\frac{\partial V_Y}{\partial Y}+V_Z\frac{\partial V_Y}{\partial Z}\right)&=-\frac{\partial P}{\partial Y}+\epsilon^2\delta^2\frac{\partial^2 V_Y}{\partial X^2}+\epsilon^2\frac{\partial^2 V_Y}{\partial Y^2}+\epsilon^4\frac{\partial^2 V_Y}{\partial Z^2},\\
    {\epsilon}Re\left(V_X\frac{\partial V_Z}{\partial X}+V_Y\frac{\partial V_Z}{\partial Y}+V_Z\frac{\partial V_Z}{\partial Z}\right)&=-\frac{\partial P}{\partial Z}+\delta^2\frac{\partial^2 V_Z}{\partial X^2}+\frac{\partial^2 V_Z}{\partial Y^2}+\epsilon^2\frac{\partial^2 V_Z}{\partial Z^2}\label{COLM-Z}.
\end{align}
\end{subequations}

As shown in figure~\ref{fig:schematic}, the fluid mechanics problem is posed on the deformed domain $\Omega_f = \{(x,y,z) \,| -w/2\le x \le +w/2,\, -h_0\le y \le u_y^0(x,z),\, 0\le z\le l\}$. We set $u_y^0=u_cU_Y^0$ and $\lambda=u_c/h_0$, where $u_c$ is the characteristic top wall deformation scale (to be determined self-consistently by solving the corresponding elasticity problem in section \ref{sec:solids}). Then, $H(X,Z)=\lambda U_Y^{0}(X,Z)$ is the dimensionless deformed top wall profile, and $\lambda$ can be interpreted as the \emph{compliance parameter} that characterizes top wall's ability to deform due to the flow beneath it. Under the lubrication approximation, which we shall now introduce, it is expected that $\lambda \ll 1/\delta$ \cite{CCSS17}. Further assuming that $\epsilon Re \ll 1$, the leading-order solution of  equations~\eqref{iNS}, subject to no-slip condition at the top and bottom walls of the channel, is
\begin{equation}\label{vzdim}
    V_Z(X,Y,Z) = \frac{1}{2}\frac{\mathrm{d}P}{\mathrm{d}Z}(Y+1)[Y-H(X,Z)]\qquad (-1 \le Y \le H).
\end{equation}
At this stage, the (non-constant) pressure gradient $\mathrm{d}P/\mathrm{d}Z <0$ remains unknown. Due to the chosen scaling (to balance conservation of mass), the velocity components $V_X$ and $V_Y$ come in at higher orders in the perturbation expansion. Systematic corrections in powers of $\epsilon$ can be obtained as regular perturbations  \cite{THFS14}, however the expansion in $\delta$ is singular \cite{D17b}. For our purposes, it suffices to note that the flow is primarily unidirectional in the $z$-direction. This is the familiar Reynolds lubrication approximation \cite{Reynolds1886,L07}. It is also useful to write down the dimensional form of equation \eqref{vzdim} for the convenience in the upcoming discussion:
\begin{equation}\label{vz}
    v_z(x,y,z) = \frac{1}{2\mu}\frac{\mathrm{d}p}{\mathrm{d}z}(y+h_0)\left[y-u_y^0(x,z)\right]\qquad \left(-h_0 \le y \le u_y^0\right).
\end{equation}

At the leading order, the balance between the fluid's shear stress and the pressure ensures the conservation of linear momentum in the $z$-direction (see equation \eqref{COLM-Z}). It is useful to rewrite the equation explicitly in the form of stresses and make dimensionless in order to determine the ordering of the forces, which reads
\begin{equation}
    -\frac{\mathcal{P}_0}{l}\frac{\partial P}{\partial Z}+\frac{\mathfrak{T}_0}{h_0}\frac{\partial \mathcal{T}_{YZ}}{\partial Y}=0.
\end{equation}
Here, we have introduced $\mathfrak{T
}_0$ as the characteristic scale for the fluid's shear stress $\mathcal{T}_{YZ}$. The force balancing requires the prefactors to be of the same order, thus $\mathfrak{T}_0=\epsilon \mathcal{P}_0$, i.e., the shear stress in the flow is much smaller than the pressure. Therefore, when dealing with the solid mechanics problem in the next section, the hydrodynamic pressure is regarded as the only applied force that causes deformation.

\section{Governing equations of the solid mechanics problem}
\label{sec:solids}

\subsection{Plane strain configuration and the thickness effect}
\label{sec:solid_scaling}
It should be clarified beforehand that we ``cut off'' the solid from the sides and only consider the deformation of the top rectangular slab, with width $w$, thickness $t$ and length $l$ with initial configuration occupying the domain $\Omega_{s0} = \{(x,y,z) \,| -w/2\le x \le +w/2,\, 0 \le y \le t,\, 0\le z\le l\}$. For convenience, we introduce the parameter $\gamma=t/h_0$ to denote the ratio of the solid thickness to the undeformed channel height and assume that $t \ll l$, equivalent to $\gamma\epsilon \ll 1$. Furthermore, the deformation is assumed to be much smaller than any dimension of the solid (i.e., small deformation gradient) so that the theory of linear elasticity applies, {i.e., $u_c\ll w$ and $u_c\ll t$. In terms of the compliance parameter $\lambda$, this assumption requires that $\lambda \ll \mathrm{max} (1/\delta,\gamma)$.} Neglecting body forces, the elastostatics equations are simply reduced to the balance between the Cauchy stresses: 
\begin{subequations}\label{equi}
\begin{align}
    \frac{\partial \sigma_{xx}}{\partial x}+\frac{\partial \sigma_{xy}}{\partial y}+\frac{\partial \sigma_{xz}}{\partial z} &=0,\\
    \frac{\partial \sigma_{xy}}{\partial x}+\frac{\partial \sigma_{yy}}{\partial y}+\frac{\partial \sigma_{yz}}{\partial z} &=0,\\
    \frac{\partial \sigma_{xz}}{\partial x}+\frac{\partial \sigma_{yz}}{\partial y}+\frac{\partial \sigma_{zz}}{\partial z} &=0,
\end{align}
\end{subequations}
where $\sigma_{xx}$, $\sigma_{yy}$, $\sigma_{zz}$, $\sigma_{xy}$, $\sigma_{xz}$ and $\sigma_{yz}$ are the six independent variables.

The next natural step is to find the primary force balancing by making the elastostatics equations dimensionless. First, the scale for $y$ should now be taken as $\gamma h_0 = t$ in order to account for the thickness effect, thus let $\mathcal{Y} = y/t$. Second, the continuity of stresses at the fluid--solid interface gives a clue of how to pick the scales for the stresses in the solid. It has been shown in section~\ref{sec:fluids} that $p \sim \mathcal{P}_0$ and $\tau_{yz} \sim \epsilon \mathcal{P}_0$. Since $\sigma_{yy}=-p$ and $\tau_{yz}=\sigma_{yz}$ at the fluid--solid interface, the scales for $\sigma_{yy}$ and $\sigma_{yz}$ should be $\mathcal{P}_0$ and $\epsilon \mathcal{P}_0$, respectively. What is more, it is believed that $\sigma_{xz}$ is negligible because $\tau_{xz}= \mathcal{O}(\epsilon\delta)$ is even smaller. {Thus, we henceforth neglect $\sigma_{xz}$ from the equilibrium equations \eqref{equi}.} Accordingly, we make the stress components of the solid dimensionless as
\begin{equation}
    \Sigma_{XX} =\frac{\sigma_{xx}}{\mathcal{D}_{xx}},\quad \Sigma_{\mathcal{Y}\mathcal{Y}}=\frac{\sigma_{yy}}{\mathcal{P}_0},\quad
    \Sigma_{ZZ}=\frac{\sigma_{zz}}{\mathcal{D}_{zz}},\quad \Sigma_{X\mathcal{Y}}=\frac{\sigma_{xy}}{\mathcal{D}_{xy}},\quad \Sigma_{\mathcal{Y}Z}=\frac{\sigma_{yz}}{\epsilon\mathcal{P}_0}.
\end{equation}
where $\mathcal{D}_{xx}$, $\mathcal{D}_{zz}$, and $\mathcal{D}_{xy}$ are the (\textit{a priori} unknown) characteristic scales for $\sigma_{xx}$, $\sigma_{zz}$ and $\sigma_{xy}$, respectively. Then the dimensionless elastostatics equations become
\begin{subequations}\label{solid_stress}
\begin{align}
    \frac{\mathcal{D}_{xx}}{w}\frac{\partial\Sigma_{XX}}{\partial X}+\frac{\mathcal{D}_{xy}}{\gamma h_0}\frac{\partial\Sigma_{X\mathcal{Y}}}{\partial \mathcal{Y}} &=0 \label{solid_x},\\
    \frac{\mathcal{D}_{xy}}{w}\frac{\partial\Sigma_{X\mathcal{Y}}}{\partial X}+\frac{\mathcal{P}_0}{\gamma h_0}\frac{\partial\Sigma_{\mathcal{YY}}}{\partial \mathcal{Y}}+\frac{\epsilon \mathcal{P}_0}{l}\frac{\partial\Sigma_{\mathcal{Y}Z}}{\partial Z} &=0 \label{solid_y},\\
     \frac{\epsilon\mathcal{P}_0}{\gamma h_0}\frac{\partial\Sigma_{\mathcal{Y}Z}}{\partial \mathcal{Y}}+\frac{\mathcal{D}_{zz}}{l}\frac{\partial\Sigma_{ZZ}}{\partial Z}&=0\label{solid_z}.
\end{align}
\end{subequations}
From equation \eqref{solid_y}, the normal stress has to be balanced with the shear stress, at the leading order in $\epsilon$, thus $\mathcal{D}_{xy}=\mathcal{P}_0/(\gamma \delta)$. Substituting this scale into equations \eqref{solid_x} and \eqref{solid_z}, we obtain $\mathcal{D}_{xx}=\mathcal{P}_0/(\gamma^2\delta^2)$ and $\mathcal{D}_{zz}=\mathcal{P}_0/\gamma$. It is important to note that $\gamma$ (i.e., the solid thickness parameter) plays an essential role in the stress distribution. Accordingly, the boundary condition at the sidewalls $x=\pm w/2$, due to the reaction between the top wall and the remaining solid, will give rise to the thickness effect.

In the present study, we are interested in the thickness range of $t\ll l$, i.e., $\gamma \ll 1/\epsilon$, thus $\gamma \delta \ll \delta/\epsilon$. In this distinguished limit, we have shown that $\sigma_{xx}$, $\sigma_{xy}$, $\sigma_{yy}$ and $\sigma_{zz}$ are the dominant stresses in the solid as $\epsilon\to0^+$. Furthermore, based on the linear constitutive relation between the stress and the linear strain, as well as the fact that the microchannel is usually prevented from displacements  in the flow-wise direction by rigid inlet and outlet connectors, we are justified in reducing the problem to a plane strain configuration with dominant linear strains $e_{xx}$, $e_{yy}$ and $e_{yz}$, in the cross-section.

For a plane strain problem, it is convenient to introduce the Airy stress function $\phi(x,y)$ that satisfies the homogeneous biharmonic equation (in dimensional form) \cite{LRK99}:
\begin{equation}\label{airy}
    \frac{\partial^4 \phi}{\partial x^4}+2\frac{\partial^4 \phi}{\partial x^2 \partial y^2}+\frac{\partial^4 \phi}{\partial y^4}=0.
\end{equation}
Then, the stresses are computed from the stress function as $\sigma_{xx}=\partial^2 \phi/\partial y^2$, $\sigma_{yy}=\partial^2 \phi/\partial x^2$ and $\sigma_{xy}=-\partial^2 \phi/\partial x\partial y$. Meanwhile,  $\sigma_{zz}=\nu(\sigma_{xx}+\sigma_{yy})$ according to the constitutive equation of linear elasticity, where $\nu$ is the Poisson ratio. Note our analysis necessitates different characteristic scales for the $y$ coordinate for the fluid and solid mechanics problems. Therefore, for consistency and convenience, we solve the solid mechanics problem in its dimensional form~\eqref{airy}.

\subsection{Large-thickness case}
\label{sec:solid_solution}

In the present study, as discussed in section~\ref{sec:intro}, we are interested in the case of thick top wall, which arises because microchannels are frequently embedded in half-space-like PDMS medium when manufactured by, e.g., replica molding \cite{SMMC11}. From the scaling analysis of equation \eqref{solid_stress} above, we learned that, as the thickness increases, $\sigma_{xy}$ decreases as $1/(\gamma\delta)$, while $\sigma_{xx}$ decreases even faster, as $1/(\gamma\delta)^2$. Thus, consider the case when the thickness is large enough, specifically $(\gamma\delta)^2 \gg 1$. Then, $\sigma_{xx}$ is much smaller than the other stresses in the solid, as well as at both sidewalls. Taking $\sigma_{xx}|_{x=\pm w/2}=0$ and assuming that the displacement at the corner is negligible, the boundary condition at $x=\pm w/2$ is reduced to that of a simple support. This result is crucial to the analysis that  follows because equation \eqref{airy} can now be solved exactly using Fourier series in the simply supported rectangular configuration. 

First, the stress at the fluid--solid interface is decomposed into Fourier series. At $y=0$, {the} normal stress in the solid should match the local hydrodynamic pressure:
\begin{equation}\label{interstr}
    \sigma_{yy}|_{y=0}=-p(z)=-p(z)\sum_{m=1}^{\infty}A_m\sin\left[\kappa_m\left(x+\frac{w}{2}\right)\right],
\end{equation}
where $\kappa_m = {m\pi}/{w}$ and $A_m=\frac{2}{m\pi}[1-(-1)^m]$. We have used the fact that the Fourier series part of \eqref{interstr} equals $1$ for $x\in(-w/2,+w/2)$. Note, however, that the series converges to $0$ at $x=\pm {w}/{2}$ because the odd extension has been used here to construct sine series, which causes the discontinuity at the two edges.

Next, the superposition principle comes into play. It is easily verified that
\begin{equation}
    \phi_m(x,y) = \sin \left[\kappa_m \left(x+\frac{w}{2}\right)\right] \left(C_1e^{\kappa_m y}+C_2e^{-\kappa_m y}+C_3 ye^{\kappa_m y}+C_4ye^{-\kappa_m y}\right)
\end{equation}
satisfies equation \eqref{airy} for any integer $m=1,2,\hdots$. The corresponding stress state is
\begin{subequations}\label{strm}
    \begin{align}
    \sigma_{xx,m}(x,y) 
        &=\sin\left[\kappa_m \left(x+\frac{w}{2}\right)\right] \big[C_1\kappa_m^2 e^{\kappa_m y}+C_2\kappa_m^2 e^{-\kappa_m y}\nonumber\\
    &\quad+C_3(2\kappa_m e^{\kappa_m y}+  e^{\kappa_m y})
    +C_4(-2\kappa_m e^{-\kappa_m y}+\kappa_m^2 y e^{-\kappa_m y})\big],\label{eq:s_xx_m}\displaybreak[3]\\
    \sigma_{yy,m}(x,y) 
        &=-\kappa_m^2 \sin\left[\kappa_m \left(x+\frac{w}{2}\right)\right] \big(C_1e^{\kappa_m y}+C_2e^{-\kappa_m y}+C_3 ye^{\kappa_m y}+C_4ye^{-\kappa_m y}\big),\label{eq:s_yy_m}\displaybreak[3]\\
    \sigma_{xy,m}(x,y) 
        &=-\kappa_m\cos\left[\kappa_m \left(x+\frac{w}{2}\right)\right] \big[C_1{\kappa_m}e^{\kappa_m y}-C_2{\kappa_m} e^{-\kappa_m y}\nonumber\\
        &\quad+C_3(e^{\kappa_m y}+{\kappa_m} y e^{\kappa_m y})+C_4(e^{-\kappa_m y}-{\kappa_m} y e^{-\kappa_m y})\big].\label{eq:s_xy_m}
    \end{align}
\end{subequations}
Four boundary conditions are needed to determine these coefficients. The stress continuity at the interface (equations \eqref{bc}$_1$ and \eqref{bc}$_2$) and the stress free conditions at the upper edge of the top wall (equations \eqref{bc}$_3$ and \eqref{bc}$_4$) require that
\begin{equation}\label{bc}
    \sigma_{yy,m}|_{y=0}=A_m\sin\left[\kappa_m \left(x+\frac{w}{2}\right)\right],\quad
    \sigma_{xy,m}|_{y=0}=0,\quad
    \sigma_{yy,m}|_{y=t}=0,\quad
    \sigma_{xy,m}|_{y=t}=0.
\end{equation}
Imposing equations \eqref{bc} on equations \eqref{strm}, we obtain
\begin{subequations}\label{coeffi}
\begin{align}
C_1 &=-\frac{A_m(1+2\beta+2\beta^2-e^{-2\beta})}{2\kappa_m^2(1+2\beta^2-\cosh{2\beta})}\label{C1},\\
C_2 &=-\frac{A_m(1+2\beta e^{-2\beta}-2\beta^2 e^{-2\beta}-e^{-2\beta})}{\kappa_m^2[(e^{-2\beta}-1)^2-4\beta^2 e^{-2\beta}]}\label{C2},\\
C_3 &=\frac{A_m(1+2\beta-e^{-2\beta})}{2\kappa_m(1+2\beta^2-\cosh{2\beta})}\label{C3},\\
C_4 &=-\frac{A_m(1+2\beta e^{-2\beta}-e^{-2\beta})}{\kappa_m[(e^{-2\beta}-1)^2-4\beta^2 e^{-2\beta}]}.\label{C4}
\end{align}
\end{subequations}
Note the coefficients $C_1$, $C_2$, $C_3$ and $C_4$ are not fixed constants but vary with $m$ and the thickness $t$ via the definition $\beta=\kappa_m t=m\pi\gamma\delta$. 
Finally, the solution to equation \eqref{airy}, as well as the three unique stress components, can be constructed by superposition:
\begin{multline}
    \phi(x,y,z) =-p(z)\sum_{m=1}^{\infty}\phi_m(x,y), \qquad
    \sigma_{xx}(x,y,z) =-p(z)\sum_{m=1}^{\infty}\sigma_{xx,m}(x,y), \\
    \sigma_{yy}(x,y,z) =-p(z)\sum_{m=1}^{\infty}\sigma_{yy,m}(x,y), \qquad
    \sigma_{xy}(x,y,z) =-p(z)\sum_{m=1}^{\infty}\sigma_{xy,m}(x,y).
\end{multline}

\subsection{Displacements and the shape of the fluid--solid interface}
\label{sec:interface}
Of course, the analysis above is only valid for small deformation gradients. In this regime, the stress--strain relations of linear elasticity \cite{LRK99} dictate that
\begin{subequations}\label{hooke}
\begin{alignat}{2}
    e_{xx,m} &=\frac{\partial u_{x,m}}{\partial x} &&= \frac{1}{\overline{E}_Y}\left(\sigma_{xx,m}-\overline{\nu}\sigma_{yy,m}\right)\label{epsx},\\
    e_{yy,m} &= \frac{\partial u_{y,m}}{\partial y} &&= \frac{1}{\overline{E}_Y}\left(\sigma_{yy,m}-\overline{\nu}\sigma_{xx,m}\right)\label{epsy},\\
    e_{xy,m} &= \frac{1}{2}\left(\frac{\partial u_{y,m}}{\partial x}+\frac{\partial u_{x,m}}{\partial y}\right) &&= \frac{1}{2G}\sigma_{xy,m},
    \label{gamxy}
\end{alignat}
\end{subequations}
where $G=E/[2(1+\nu)]$ is the shear modulus of elasticity, and $\overline{E}_Y$ and $\overline{\nu}$ are related to the Young's modulus $E_Y$ and the Poisson's ratio $\nu$ by $\overline{E}_Y={E_Y}/(1-\nu^2)$ and  $\overline{\nu}={\nu}/(1-\nu)$, respectively, because of the plane strain configuration considered herein \cite{LRK99}.

Integrating equations \eqref{epsx} and \eqref{epsy}, $u_{x,m}$ and $u_{y,m}$ are, respectively,
\begin{subequations}\begin{align}
\label{uxm}
    u_{x,m}(x,y) &= -\frac{1}{\overline{E}_Y}\cos\left[\kappa_m \left(x+\frac{w}{2}\right)\right] \Big\{[-2C_4+\kappa_m(1+\overline{\nu})(C_2+C_4y)]e^{-\kappa_m y}\nonumber\\
    &\phantom{=}+[2C_3+\kappa_m(1+\overline{\nu})(C_1+C_3 y)]e^{\kappa_m y}\Big\}+f_2(y),\\
\label{uym}
    u_{y,m}(x,y) &= \frac{1}{\overline{E}_Y}
    \sin\left[\kappa_m \left(x+\frac{w}{2}\right)\right] \Big\{[C_4(1-\overline{\nu})+\kappa_m(1+\overline{\nu})(C_2+C_4y)]e^{-\kappa_m y}\nonumber\\
    &\phantom{=}-[-C_3(1-\overline{\nu})+\kappa_m(1+\overline{\nu})(C_1+C_3 y)]e^{\kappa_m y}\Big\}+f_1(x),
\end{align}\label{eq:uxym}\end{subequations}
where  $f_1(x)$ and $f_2(y)$ are arbitrary functions of integration. Substituting equations \eqref{uxm} and \eqref{uym} into equation \eqref{gamxy}, we find
\begin{equation}\label{f1f2}
    \frac{\partial f_1}{\partial x}+\frac{\partial f_2}{\partial y}=0.
\end{equation}
From equation \eqref{f1f2}, it is easily concluded that both $f_1$ and $f_2$ should be constants. Since it is assumed that there are no vertical displacement at $x=\pm {w}/{2}$, $f_1=0$. According to the symmetry of the problem, i.e., $({\partial u_x}/{\partial x})|_{x=0}=0$, $f_2=0$ as well.

Finally, the displacements are obtained by summing up all the $u_{x,m}$ and $u_{y,m}$ terms from equations~\eqref{eq:uxym}:
\begin{equation}
    u_x(x,y,z) = -p(z)\sum_{m=1}^{\infty}u_{x,m}(x,y),\qquad
    u_y(x,y,z) = -p(z)\sum_{m=1}^{\infty}u_{y,m}(x,y).
\label{disp}    
\end{equation}
To obtain the fluid--solid interface deflection profile, $u_y^0(x,y)$, as well as the shape of the whole deformed cross-section, we simply take $y=0$ in equations \eqref{disp}. In order to evaluate the Fourier series numerically and generate the plots herein, we find that keeping 50 terms in the sum is sufficient.

\begin{figure}
    \centering
    \includegraphics[width=\textwidth]{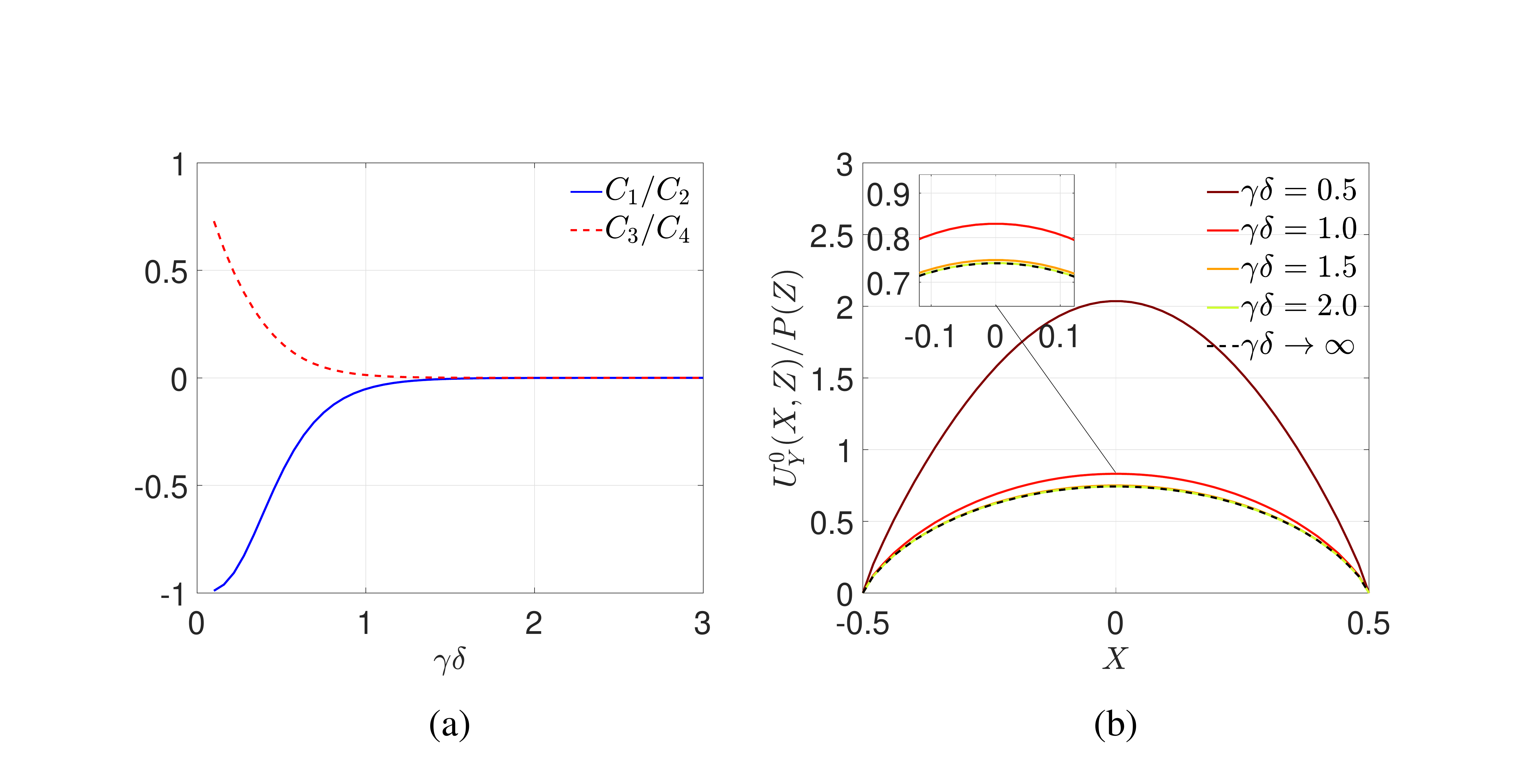}
    \caption{(a) Comparison of the coefficients from equations \eqref{coeffi}. (b) The fluid--solid interface deflection profile from equations \eqref{disp}$_2$, \eqref{uyinf} and \eqref{uyinf_dimless} for different thickness-to-width ratios $\gamma\delta=t/w$. The magnified plot was generated using the script from \cite{mag_plot}.}
    \label{fig:c_uy}
\end{figure}

However, in the large thickness case of interest herein, the above results can be further simplified because $C_1$ and $C_3$ are small compared to $C_2$ and $C_4$, respectively. As shown in figure \ref{fig:c_uy}, the dimensionless ratios $C_1/C_2$ and $C_3/C_4$ decrease very quickly with $\gamma\delta$. Specifically, for $t/w \gtrsim 1$ (say, $t \simeq 1.5 w$), $C_1$ and $C_3$ are negligible compared to $C_2$ and $C_4$. In this case, we can simply regard the stress-free boundary conditions in equations \eqref{bc}$_3$ and \eqref{bc}$_4$ as being satisfied at $y=\infty$ instead of $y=t$; hence, $C_1=C_3=0$, $C_2=-A_m/\kappa_m^2$ and $C_4=-A_m/\kappa_m$. Then, the vertical displacement at the fluid--solid interface reduces to  
\begin{equation}\label{uyinf}
    u_y^0(x,z)=\frac{p(z)}{\overline{E}_Y}\sum_{m=1}^{\infty}\frac{2A_m}{\kappa_m}\sin\left[\kappa_m\left(x+\frac{w}{2}\right)\right] \qquad (\gamma^2\delta^2=t^2/w^2\to\infty).
\end{equation}
The panel of figure \ref{fig:c_uy} supports our observations. The interface profiles for the cases of $\gamma\delta=1.5$ and $\gamma\delta=2.0$ coincide with the curve predicted by equation \eqref{uyinf}.

It is easy to rewrite equation \eqref{uyinf} in dimensionless form as
\begin{equation}\label{uyinf_dimless}
    \frac{u_c\overline{E}_Y}{w \mathcal{P}_0}U_Y^0(X,Z)=P(Z)\sum_{m=1}^{\infty}\frac{2A_m}{m\pi}\sin\left[m\pi\left(X+\frac{1}{2}\right)\right]=P(Z)\mathfrak{G}(X),
\end{equation}
where, for convenience, we have denoted by $\mathfrak{G}$ the function of $X$ defined by the Fourier series. Now, the natural deformation scale is clearly $u_c=w\mathcal{P}_0/\overline{E}_Y$, so that we can set the prefactor on the left-hand side of equation \eqref{uyinf_dimless} to unity. This scale is similar to the one used in \cite{GEGJ06}, where it was assumed $\langle u_y \rangle /w\sim p/E_Y$. Note, however, that our analysis shows that $\overline{E}_Y=E_Y/(1-\nu^2)$ must be used in the deformation scale \emph{instead} of $E_Y$ because the top wall in a long, shallow microchannel is in a plane strain configuration. Then, $\lambda$ in equation \eqref{vzdim} is finally determined to be  $\lambda=u_c/h_0=\mathcal{P}_0/(\overline{E}_Y\delta)$.

We can see from equation \eqref{uyinf_dimless} that the interface deflection profiles at different $Z$ coordinates have the same shape, denoted by $\mathfrak{G}(X)$. It is easy to compute the maximum and average displacement at the interface from equation \eqref{uyinf_dimless}:
\begin{subequations}\begin{align}
    \max_X U_Y^0 &= \mathfrak{G}(0)P(Z) \approx 0.7426 P(Z),\\
    \langle U_Y^0 \rangle &= P(Z)\int_{-1/2}^{+1/2}\mathfrak{G}(X) \mathrm{d} X \approx 0.5427P(Z) \approx 0.7311\max_X U_Y^0.
\end{align}\label{uave_umax}\end{subequations}
Note the prefactors here are different from previous studies, which either assumed a parabolic deformation profile of the fluid--solid interface, in which case $\langle U_Y^0 \rangle = (2/3) \max_X U_Y^0$ \cite{OYE13}, or obtained a quartic profile from plate theory with clamped boundary condition, in which case $\langle U_Y^0 \rangle = (8/15) \max_X U_Y^0$ \cite{CCSS17}.

Observe that the simple support does not restrict the horizontal displacement (see equation \eqref{disp}$_1$). Denoting the horizontal displacement at the fluid--solid interface by $u_x^0$, we can express this displacement in the large-thickness case as
\begin{equation}\label{uxinf_dimless}
     u_x^0(x,z)=\frac{p(z)(1-\overline{\nu})}{\overline{E}_Y}\sum_{m=1}^{\infty}\frac{A_m}{\kappa_m}\cos\left[\kappa_m\left(x+\frac{w}{2}\right)\right] \qquad (\gamma^2\delta^2=t^2/w^2\to\infty).
\end{equation}
Given that the typical Poisson ratio of PDMS is $\nu \approx 0.4$ to 0.5 \cite{JMTT14}, our theory predicts $u_x^0 \ll u_y^0$ since the ratio of the maximum value of these two displacements, from equations \eqref{uyinf_dimless} and \eqref{uxinf_dimless}, is only $(1-\overline{\nu})/2=(1-2\nu)/[2(1-\nu)]$. This result is consistent with experimental observations \cite{GEGJ06}. Interestingly, if we take the material of the top wall to be strictly incompressible with $\nu=0.5$, the $u_x^0$ is exactly zero, even though $u_x\ne0$ for $y>0$. The latter is not important in the context of the present study because we focus on the fluid domain's shape inside the microchannel.

\subsection{Summary and discussion of the solid mechanics results}

To summarize, we have derived a mathematical expression for the fluid--solid interface deflection curve for the large thickness case. It should be clarified again that the thickness is considered ``large'' specifically when $1/(\gamma\delta)^2=(w/t)^2\ll1$. In this distinguished limit, we have shown that the top wall can be considered as a simply supported rectangle subject to uniform pressure at the bottom. Importantly, note the present thickness range includes but is wider than $w/t\ll1$. For some cases with $w/t\simeq1$, asymptotically, we can still satisfy $(w/t)^2\ll1$.

An important case is the static response of microchannels with thinner top wall, such as a plate-like structure. However, our  configuration, based on the fabrication methods from, e.g., \cite{GEGJ06} and depicted in figure \ref{fig:schematic}, is different from previous studies \cite{RS16, OYE13, CCSS17, SC18, BGB18, MY19} in which the top wall is modelled as a clamped membrane or plate with small thickness $t\lesssim w$. Consequently, $1/(\gamma\delta)^2\ll 1$ is not satisfied and the analysis above is inapplicable. Nevertheless, the plane strain assumption is still valid. After analyzing the resultant forces and moments in the solid, we hypothesize that in this case, at each fixed-$z$ cross-section, the top wall behaves like a simply supported beam with tension. However, it is hard to find the exact solutions for the biharmonic equation \eqref{airy} under such boundary conditions. Therefore, we use beam theory to obtain the fluid--solid interface profile and the hydrodynamic resistance. These details are provided in the Appendix.

\section{Hydrodynamic resistance of the compliant channel}
\label{sec:resistance}
Having solved for the leading-order velocity profile in section \ref{sec:fluids} and the cross-sectional shape of the fluid--solid interface in section \ref{sec:interface}, we are now in a position to solve the coupled fluid--structure interaction problem. Specifically, in the microfluidics context, of greatest interest is the \emph{hydrodynamic resistance}, which characterizes the required pressure drop (i.e., force) to maintain steady flow at a given volumetric flow rate \cite{B08}. For fixed cross-sectional shapes, this quantity can be characterized for any number of shapes \cite{MOB05} using the ability to solve the Stokes equations for $Re=0$ \cite[section 2-5]{HB83}.

In a compliant channel, however,  and the specific shape of the deformation profile in cross-section depends on the pressure itself. This results in a nonlinear relationship between the pressure drop and flow rate. We can determine this relation by directly calculating the flow rate under the deformed cross-section. On using equation \eqref{vz} for $v_z(x,y,z)$, we obtain:
\begin{equation}\label{qpdiff}
    q = \int_{-w/2}^{+w/2}\int_{-h_0}^{u_y^0(x,z)}v_z(x,y,z)\,\mathrm{d}y\,\mathrm{d}x
    = -\frac{1}{12\mu}\frac{\mathrm{d}p}{\mathrm{d}z}\int_{-w/2}^{+w/2}[h_0+u_y^0(x,z)]^3 \,\mathrm{d}x,
\end{equation}
where $u_y^0(x,z)$ is given in equation \eqref{uyinf} for the large-thickness case, and only the axial velocity component contributes to the flow rate at the leading order in the assumed small parameters. For a scenario with constant flow rate, equation \eqref{qpdiff} is a first-order differential equation for $p(z)$ given $q=const.$, which can be solved by assuming the outlet pressure sets the gauge, i.e., $p(l)=0$.

In the large-thickness case, the self-similarity of the fluid--solid interface deflection profile makes it easy to solve equation \eqref{qpdiff} by separation of variables, yielding an implicit relation for the hydrodynamic pressure:
\begin{equation}\label{qp}
    q = \frac{wh_0^3p(z)}{12\mu(l-z)}\left[1+\left(\frac{w}{\overline{E}_Yh_0}\right)S_1p(z)+\left(\frac{w}{\overline{E}_Yh_0}\right)^2 S_2p(z)^2+\left(\frac{w}{\overline{E}_Yh_0}\right)^3 S_3p(z)^3\right],
\end{equation}
where
\begin{subequations}\begin{alignat}{2}
    S_1 &= \frac{3}{2}\int_{-1/2}^{+1/2} \mathfrak{G}(X)\,\mathrm{d}X &&\approx 0.8139,\\
    S_2 &= \int_{-1/2}^{+1/2} \mathfrak{G}^2(X)\,\mathrm{d}X &&\approx0.3333,\\
    S_3 &= \frac{1}{4}\int_{-1/2}^{+1/2} \mathfrak{G}^3(X)\,\mathrm{d}X &&\approx0.05396,
\end{alignat}\label{eq:S123}\end{subequations}
and $\mathfrak{G}(X)$ is the self-similar deflection profile shared by every cross-section given in equation \eqref{uyinf_dimless}. The integrals in equations \eqref{eq:S123} are computed numerically using the trapezoidal with respect to 100 evenly-space integration points on $X\in[-1/2, +1/2]$. Observe that, while equation~\eqref{qp} has the same general structure (as already expected from \cite{RK72}) as that arising from theories based on plate-like elastic top walls \cite{CCSS17}, the prefactors $S_{1,2,3}$ related to the cross-sectional shape of the fluid--solid interface are  \emph{larger} by orders of magnitude.

Equation \eqref{qp} can also be made dimensionless in the flow-rate-controlled regime as
\begin{equation}\label{qpdim}
    Q = \frac{P(Z)}{12(1-Z)}\Big[1+S_1\lambda P(Z)+S_2\lambda^2P^2(Z)+S_3\lambda^3P^3(Z)\Big].
\end{equation}
Recall that we defined $\lambda=u_c/h_0=\mathcal{P}_0/(\overline{E}_Y\delta)$ so that $H(X,Z)=\lambda U_Y^0(X,Z)$. The dimensionless flow rate--total pressure drop relation is obtained by taking $Z=0$ in equation \eqref{qpdim}. As the top wall deformation increases, i.e., for larger $\lambda$, the nonlinearity in the relation becomes more pronounced. 

Finally, taking $z=0$, the relation between the total pressure drop $\Delta p$ and the volumetric flow rate $q$ is obtained from \eqref{qp}:
\begin{equation}\label{qdpthick}
    q = \frac{wh_0^3\Delta p}{12\mu l}\left[1+\left(\frac{w}{\overline{E}_Yh_0}\right)S_1\Delta p+\left(\frac{w}{\overline{E}_Yh_0}\right)^2 S_2(\Delta p)^2+\left(\frac{w}{\overline{E}_Yh_0}\right)^3 S_3(\Delta p)^3\right].
\end{equation}
The important message is that with the consideration of the fluid--structure interaction in the microchannel, the flow rate and pressure drop relation deviates from the classic Poiseuille's law, which for a rectangular channel is $q = wh_0^3\Delta p/(12\mu l)$ (neglecting drag from the lateral sidewalls) \cite{B08}, and displays nonlinearity. Although $S_1 > S_2 > S_3$, it is important to emphasize that equation \eqref{qdpthick} is simply a polynomial, and \emph{not} a perturbation series, in $\Delta p$.

\section{Results, discussion and illustrated examples}
\label{sec:results}

At the leading order in the small parameter $\epsilon$, we have reduced the 3D deformation of the top wall to a 2D problem by considering each cross-section (in the $(x,y)$ plane) as independent. Such decoupling is a natural consequence of the long and slender geometry and has also been shown asymptotically by Christov \textit{et al.}\ \cite{CCSS17} and numerically by Chakraborty \textit{et al.}\ \cite{CPFY12}. Based on this idea, various models, either with or without fitting parameters, have been put forward to account for the nonlinear flow rate--pressure drop relation. The very first one was from Gervais \textit{et al.}\ \cite{GEGJ06} in the form of an implicit relation for the hydrodynamic pressure $p(z)$:
\begin{equation}\label{gervais}
    q = \frac{h_0^4E_Y}{48\alpha\mu(l-z)}\left\{\left[1+\alpha\frac{p(z)w}{E_Y h_0}\right]^4-1\right\},
\end{equation}
where $\alpha$ came from the assumption $\langle u_y\rangle /h_0=\alpha p(z)/(E_Y \delta)$ for the thick top wall and has to be determined by fitting to experiments. Here $\langle u_y\rangle$ is the average interface deflection at each fixed-$z$ cross-section. Even though this model has been employed in a lot of later works \cite{CTS12,HUZK09}, the unknown fitting parameter $\alpha$ is one of the biggest drawbacks. More recent work has focused on eliminating the fitting parameter, specifically for thinner top walls, plate theory \cite{CCSS17,SC18} or engineering pressure--displacement models \cite{S11} can be used \cite{CCSS17,SC18} to obtain the hydrodynamic resistance in the deformed microchannel.

Importantly, in the present study, our emphasis is on the thick top wall case, therefore our scaling is different from \cite{CCSS17,SC18,RDC17}, wherein $\lambda = 12(w/t)^3\mathcal{P}_0/(\overline{E}_Y\delta)$ was used. A prefactor $ \propto (w/t)^3$ shows up in $\lambda$ when the top wall is plate-like, i.e., $t\lesssim w$, which is a consequence of the assumed bending-dominated regime (and results in a very large value of $\lambda$ for thick top walls). Nevertheless, these theories are self-consistent in that the coefficients in the flow rate--pressure drop relation (i.e., the counterpart of $S_1$, $S_2$ and $S_3$ in equation \eqref{qpdim} above) become much smaller to balance the large values of $\lambda$. However, in the large-thickness case, we have already shown that $\sigma_{xx}\sim 1/{(\gamma\delta)}^2$, meaning that the bending moment in the solid is actually negligible. This observation clearly shows that different solid deformation mechanisms are involved during FSI in microchannels with large versus small top wall thicknesses (compared to the width).

\begin{figure}
    \centering
    \begin{subfigure}{0.4\textwidth}
        \includegraphics[width=\linewidth]{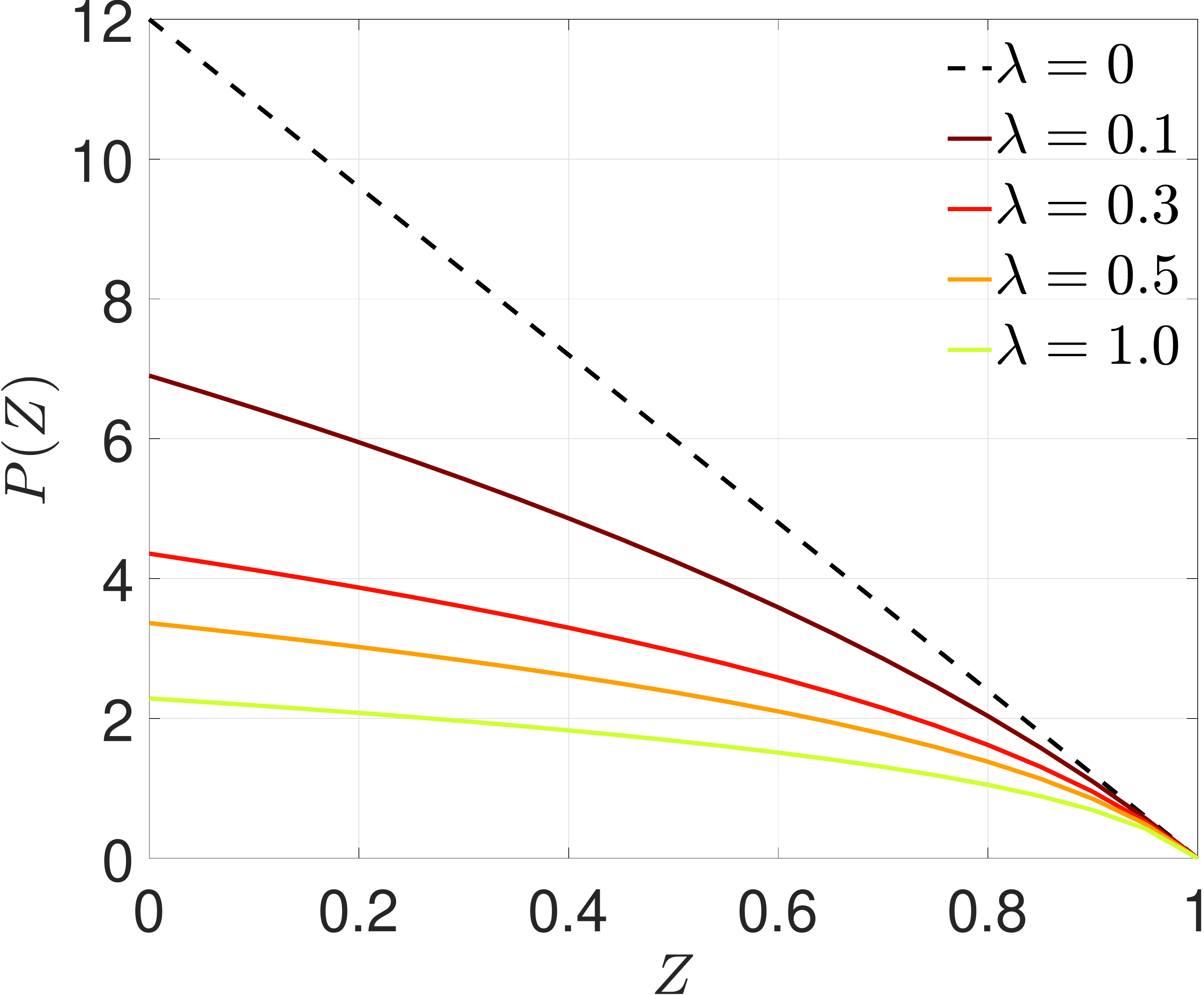}
        \caption{}
        \label{fig:PZ}
    \end{subfigure}
    \hfill
    \begin{subfigure}{0.55\textwidth}
        \includegraphics[width=\linewidth]{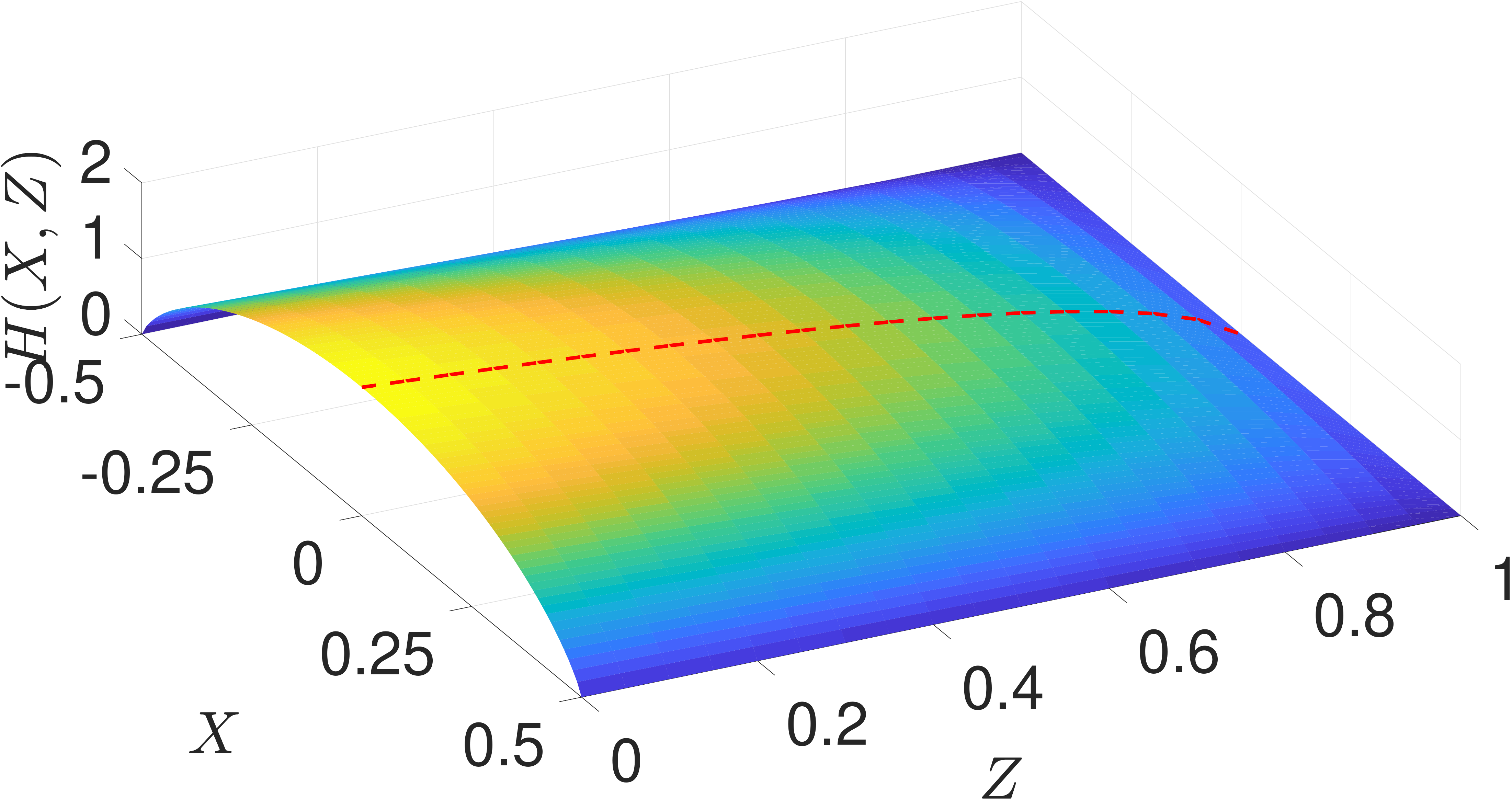}
        \caption{}
        \label{fig:surf}
    \end{subfigure}
    \caption{(a) Pressure $P$ as a function of the axial coordinate $Z$, computed by inverting equation \eqref{qpdim}, for different values of $\lambda$ with $Q=1$. (b) The deformed fluid--solid interface, $H(X,Z)=\lambda U_Y^0$, is computed from equation \eqref{uyinf_dimless} with $P(Z)$ having been obtained from \eqref{qpdim}, for $Q=1$ and $\lambda=1$. The red dashed curve represents the maximum cross-sectional deflection of the interface.}
    \label{fig:PZ_surf}
\end{figure}

\begin{figure}
    \centering
    \begin{subfigure}{0.425\textwidth}
        \includegraphics[width=\linewidth]{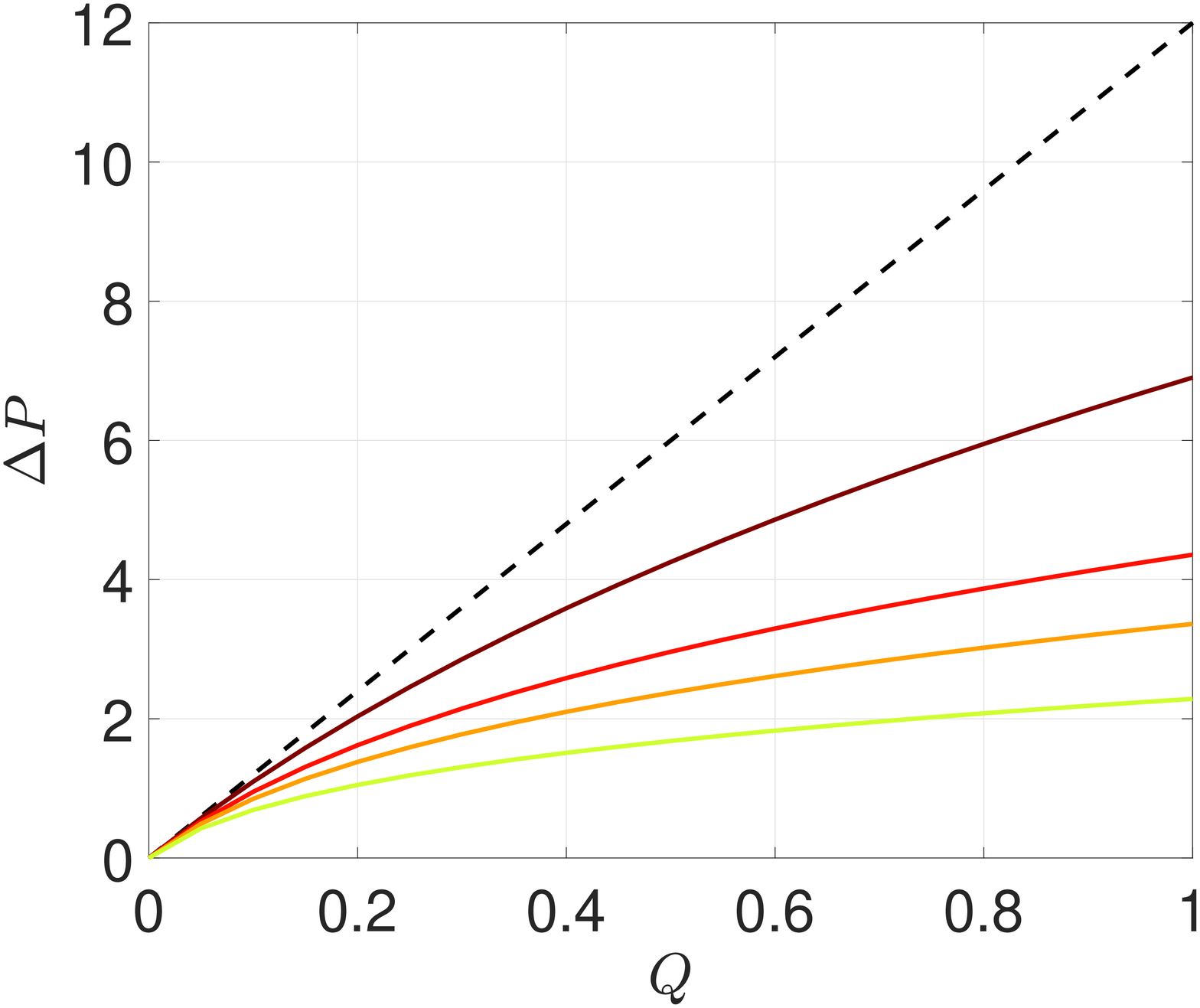}
        \caption{}
        \label{fig:QdP}
    \end{subfigure}
    \hfill
    \begin{subfigure}{0.525\textwidth}
        \includegraphics[width=\linewidth]{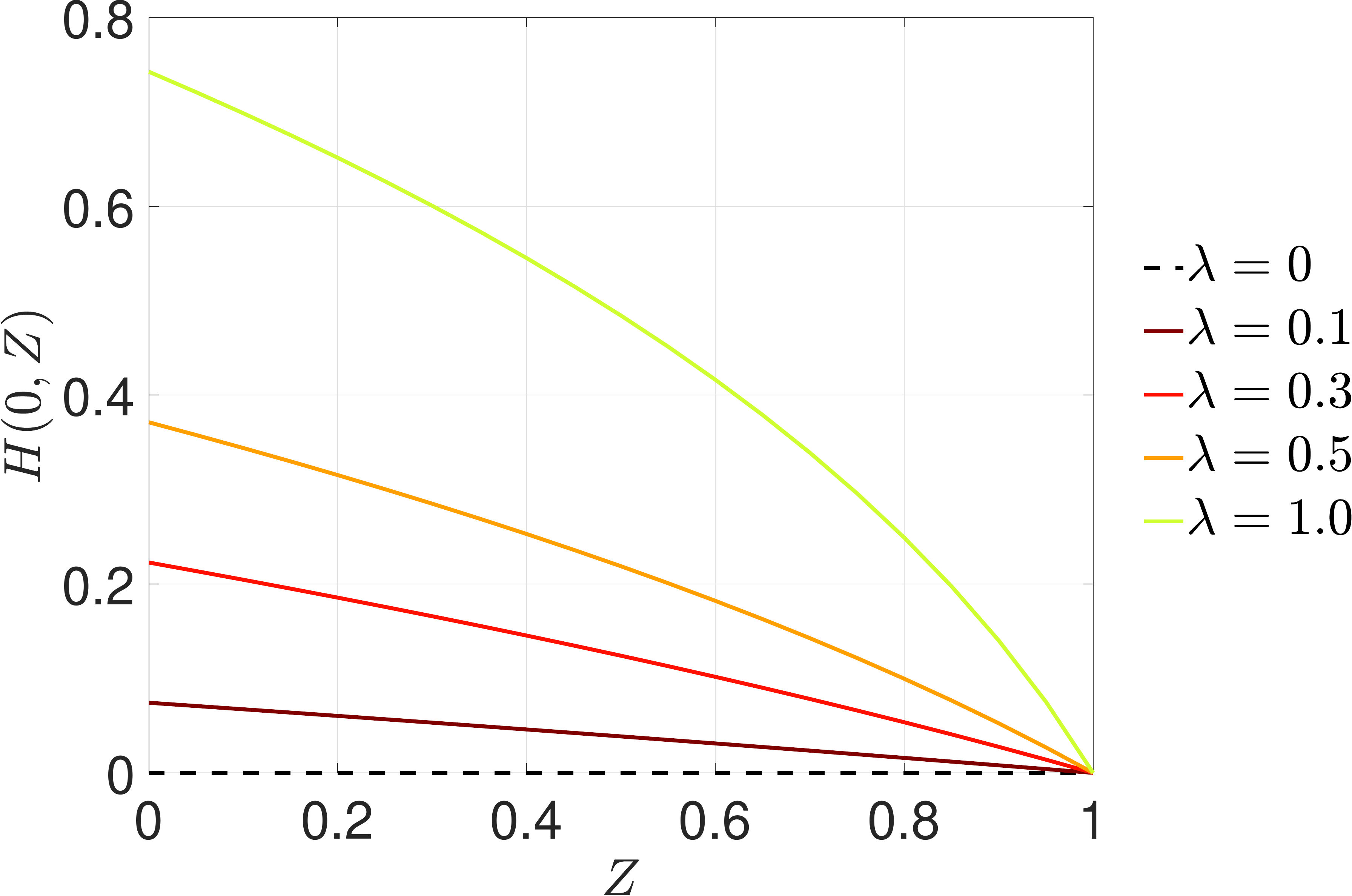}
        \caption{}
        \label{fig:DPUY}
    \end{subfigure}
    \caption{(a) Flow-rate-controlled regime: pressure drop across channel, computed using equation \eqref{qpdim}, as a function of $Q$, for different values of the compliance parameter $\lambda$. (b) Pressure-drop-controlled regime: The maximum (across the cross-section) interface deflection of the channel top wall as a function of the flow-wise coordinate $Z$ with $\Delta P=1$; $Q$ for different value of $\lambda$ is computed via equation \eqref{qpdim} evaluated at $Z=1$, then $P(Z)$ is obtained by inverting the same equation. Substituting $P(Z)$ into equation \eqref{uyinf_dimless}, $U_Y^0$ is found, from which $H(X,Z)=\lambda U_Y^0$ and $\max_{X} H(X,Z) = H(0,Z)$ are calculated and plotted.}
    \label{fig:QdP_dPUY}
\end{figure}

Next, we will give a systematic discussion of the predictions of our FSI theory for microchannels with thick top walls. At steady state, the flow rate $q=const.$, and each cross-section will inflate under the local pressure $p(z)$. The increase in area reduces the local fluid velocity, which introduces a non-constant pressure gradient along the flow-wise $z$-direction. As shown in figure \ref{fig:PZ} for $\lambda=0$ (i.e., when the channel is rigid), the pressure decreases linearly from the inlet to the outlet, and $\mathrm{d}P/\mathrm{d}Z$ is a constant in this case. However, as the top wall becomes softer, with the increase of $\lambda$, the $P(Z)$ profile deviates further from the linear profile, and $\mathrm{d}P/\mathrm{d}Z$ is a decreasing function of $Z$. Accordingly, based on equation \eqref{uyinf_dimless}, the maximum deflection at the fluid--solid interface, $H(0,Z)$, is expected to be concave, as illustrated in figures \ref{fig:surf} and \ref{fig:DPUY}.

Equation \eqref{qpdim} is applicable in both the steady-state flow-rate-controlled and pressure-drop-controlled flows. As shown in figure \ref{fig:QdP}, with controlled flow rate, the total pressure drop is a linear function of flow rate for a rigid channel but a nonlinear function for a soft channel. Furthermore, the pressure drop decreases as the compliance of the channel increases because, under a fixed flow rate, the softer channel will deform more to reduce the flow velocity, and therefore, the pressure losses due to viscosity at each cross-section. In turn, for a pressure-drop controlled flow, the softer channel will allow a higher flow rate, as well as a larger deflection. As shown in figure \ref{fig:DPUY} the maximum deflection at the fluid--solid interface also increases with $\lambda$, for a given pressure drop.

Next we compare our theory with previous experimental studies from the literature, namely \cite{GEGJ06,RDC17}. Apart from the model \eqref{gervais} proposed in \cite{GEGJ06}, Gervais \textit{et al.}\ performed experiment with microchannels with two different Young's moduli and two different widths. The important parameters are summarized in table \ref{tab:GEGJ}. Note the thickness for experiment was reported to be larger than 6 mm, but it was numerically shown that 2 mm was thick enough for a sufficiently accurate comparison. Moreover, the undeformed height, $h_0$, for the case GEGJ 4 is corrected to 30 $\mu$m instead of the reported 26 $\mu$m based on the value of $\alpha$.  We compute $1/(\gamma\delta)^2$ in the last column of table \ref{tab:GEGJ} and show that our theory is applicable to all the four cases because $1/(\gamma\delta)^2 \ll 1$ for all data sets. It is important to note that previous theory of microchannel FSI \cite{CCSS17,SC18} is not applicable, even as an approximation, to any of these cases.

\begin{table}
\centering
    \begin{tabular}{c c c c c c c c c c}
        \hline
        Case & $h_0$ & $w$ & $l$ & $t$ & $E_Y$ & $\delta$ & $\epsilon$ & $\gamma$ & $1/(\gamma\delta)^2$ \\
        & [$\mu$m] & [$\mu$m] & [cm] & [mm] & [MPa] & [\,--\,] & [\,--\,]  & [\,--\,]  & [\,--\,] \\
        \hline
        GEGJ 1 ($\blacktriangle$) & 26 & 250 & 1 & 2 & 2.2 & 0.1040 & 0.0026 & 76.92 & 0.0156 \\
        GEGJ 2 ($\blacksquare$) & 30 & 500 & 1 & 2 & 2.2 & 0.0600 & 0.0030 & 66.67 & 0.0625 \\
        GEGJ 3 ({\color{red}$\blacktriangle$}) & 26 & 250 & 1 & 2 & 1.1 & 0.1040 & 0.0026 & 76.92 & 0.0156 \\
        GEGJ 4 ({\color{red}$\blacksquare$}) & 30 & 500 & 1 & 2 & 1.1 & 0.0600 & 0.0030 & 66.67 & 0.0625 \\
        \hline
    \end{tabular}
    \caption{Values of physical parameters used in the experiments of Gervais \textit{et al.}\  \cite{GEGJ06}, where the Poisson ratio is $\nu = 0.5$, and the fluid's viscosity is $\mu = 0.001$ Pa$\cdot$s for all the cases.}
    \label{tab:GEGJ}
\end{table}

\begin{figure}
    \centering
    \includegraphics[width=0.75\textwidth]{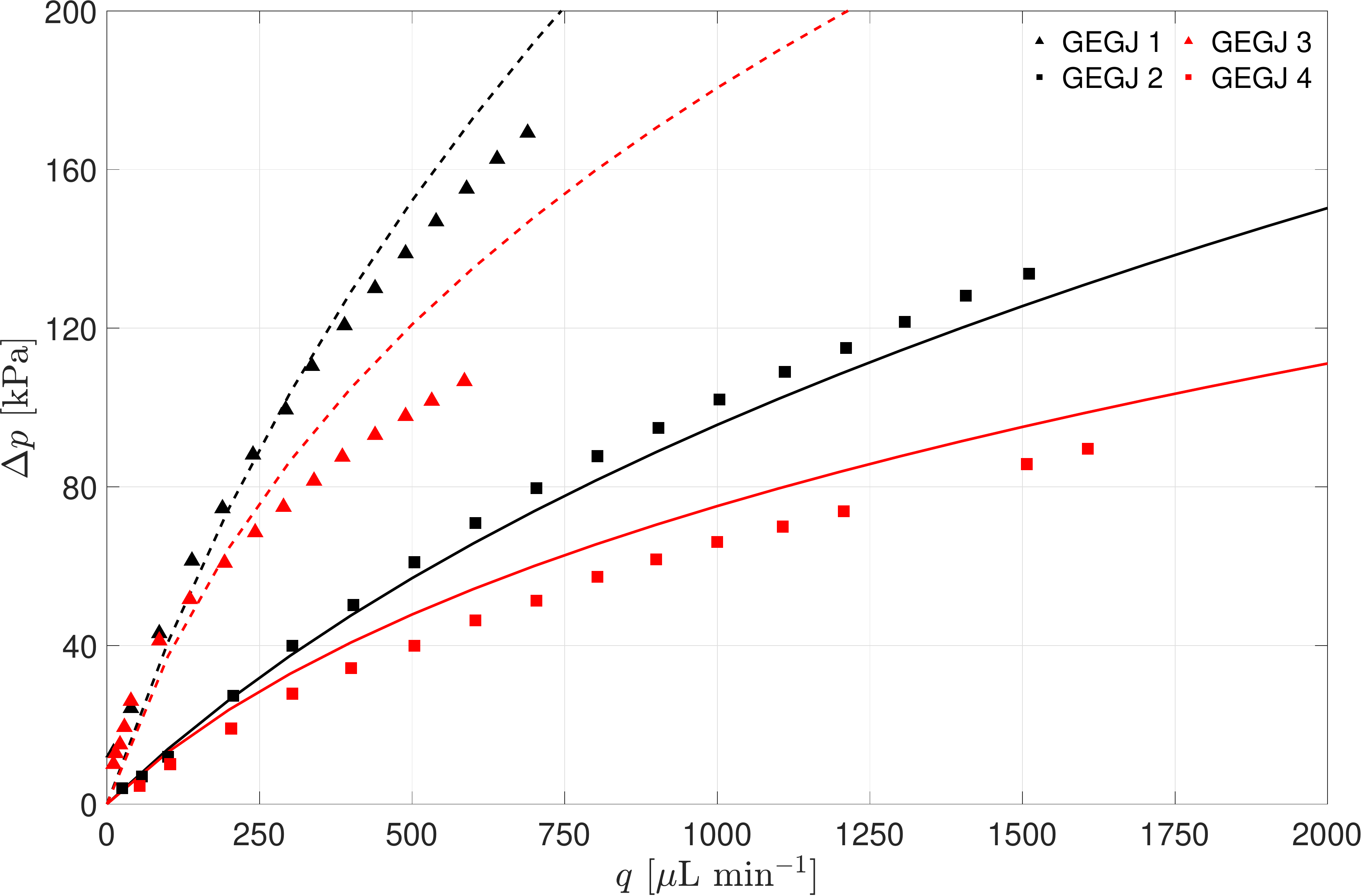}
    \caption{Comparison between our theory and the experimental data from \cite{GEGJ06} for the pressure drop $\Delta p$ as a function of the flow rate $q$. The symbols represent the experimental data while the curves are the prediction from equation \eqref{qdpthick}, without any fitting parameters. The black (dark) dashed and solid curves correspond to cases GEGJ 1 and GEGJ 2, respectively, while the red (light) curves correspond to GEGJ 3 and GEGJ 4, as described in table~\ref{tab:GEGJ}.}
    \label{fig:GEGJ_qdp}
\end{figure}

\begin{figure}
    \centering
    \includegraphics[width=0.75\textwidth]{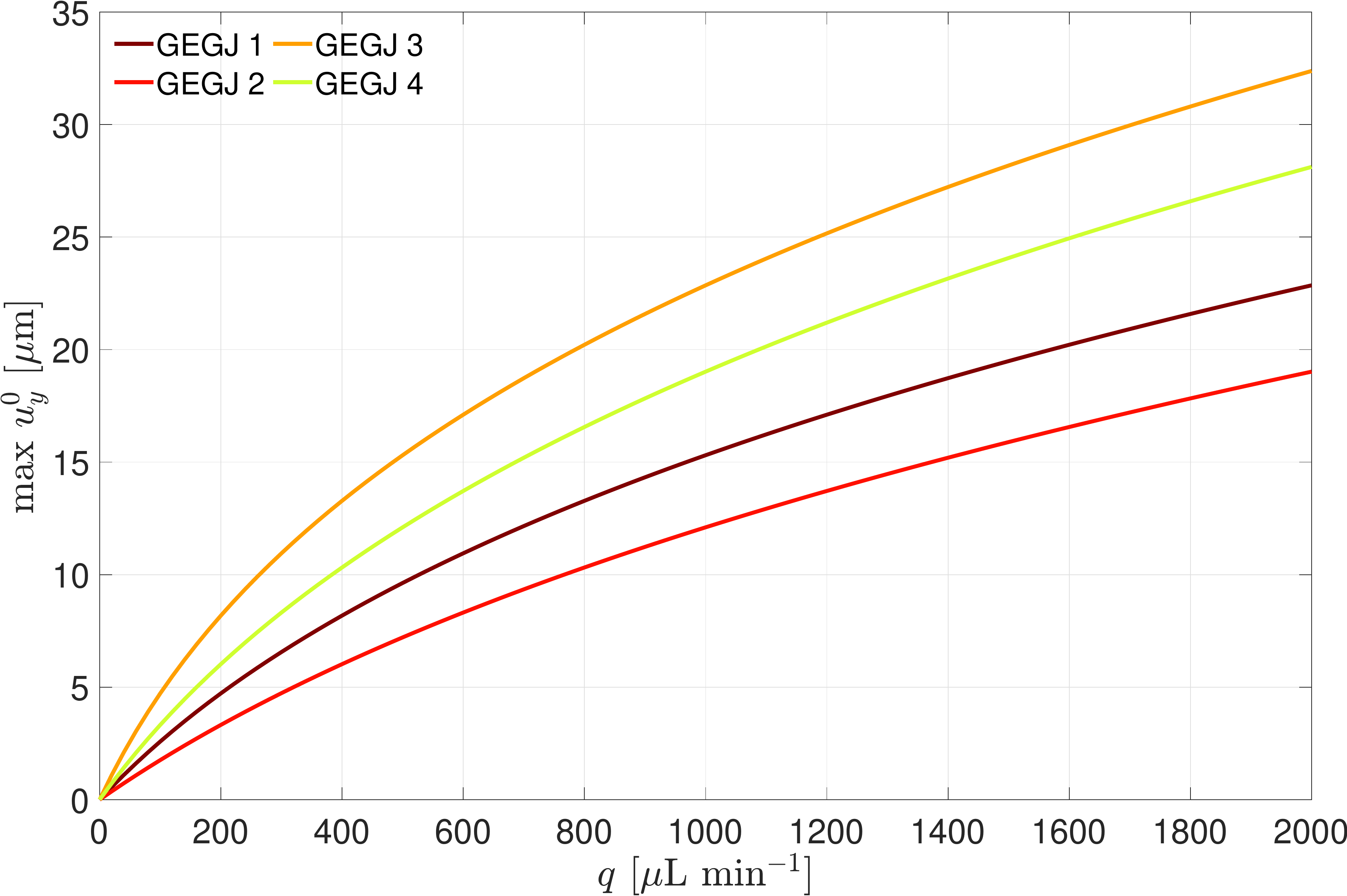}
    \caption{The maximum vertical displacement of the fluid--solid interface $\max u_y^0$ as a function of the flow rate $q$ for the cases in table \ref{tab:GEGJ}. The pressure drop is first computed by equation \eqref{qdpthick} and then substituted into equation \eqref{uyinf} to obtain  $\max u_y^0$ and plot it.}
    \label{fig:GEGJ_quymax}
\end{figure}

In figure \ref{fig:GEGJ_qdp}, the flow rate--pressure drop relation curves predicted by equation \eqref{qdpthick} are shown to agree favorably with the experiments. {The corresponding predicted maximum displacement at the interface is shown in figure \ref{fig:GEGJ_quymax}. Although some deviations are observed, it is not easy for us to provide a definite reason as to why, due to the lack of information about experimental sources of error in \cite{GEGJ06}. In the cases GEGJ 2 and 4, there exists an almost constant shift from the experiment, which could be  systematic error. For the cases GEGJ 1 and GEGJ 3, the pressure drops predicted by the theory at higher flow rates are larger than the experiments, which would indicate that the theory underestimates the channel deformation at higher flow rates. It is also relevant to note that the worst agreement in figure~\ref{fig:GEGJ_qdp} is for case GEGJ 3, which exhibits the largest deformation in figure~\ref{fig:GEGJ_quymax}. Overall, the largest source of uncertainty, however, is the measurement of the undeformed channel height $h_0$. Indeed, one reason we have not included the maximum deformation data from \cite{GEGJ06} in figure \ref{fig:GEGJ_quymax} is that the error bars are too large to make a meaningful comparison. As we show in the next comparison with the experiments from \cite{RDC17}, a small uncertainty in $h_0$ can lead to a large effect on the predicted pressure drop.}

A further, quantitative, comparison between our theory and the fitting model \eqref{gervais} can be achieved by computing the values of the statistical coefficient of determination $R^2$ via least squares \cite{WellsR2_other} for each model, as shown in table \ref{tab:GEGJ_R2}. Unsurprisingly, the $R^2$ values of the model \eqref{gervais} are closer to 1 than those of the present theory because it is a one-parameter best-fit of the experimental data. Nevertheless, the present theory, without any fitting parameters, also give values of $R^2 \approx 1$, which means that the present fitting-parameter-free theory can capture the physics of the problem as accurately as a fitting model.

\begin{table}
    \centering
    \begin{tabular}{c c c c c}
        \hline
       Case  &  GEGJ 1 & GEGJ 2 & GEGJ 3 & GEGJ 4 \\
       \hline
    Present theory & 0.9633 & 0.9859 & 0.8904 & 0.9224 \\
    GEGJ fit & 0.9904 & 0.9988 & 0.9792 & 0.9920\\
    \hline
    \end{tabular}
    \caption{The comparison of the values of $R^2$ between the present theory, equation \eqref{qdpthick}, and the model in \cite{GEGJ06}, denoted GEGJ fit (see equation \eqref{gervais}). The value of the fitting parameter $\alpha$ for each case is available in \cite{GEGJ06}. }
    \label{tab:GEGJ_R2}
\end{table}

\begin{figure}
    \centering
    \includegraphics[width=0.75\textwidth]{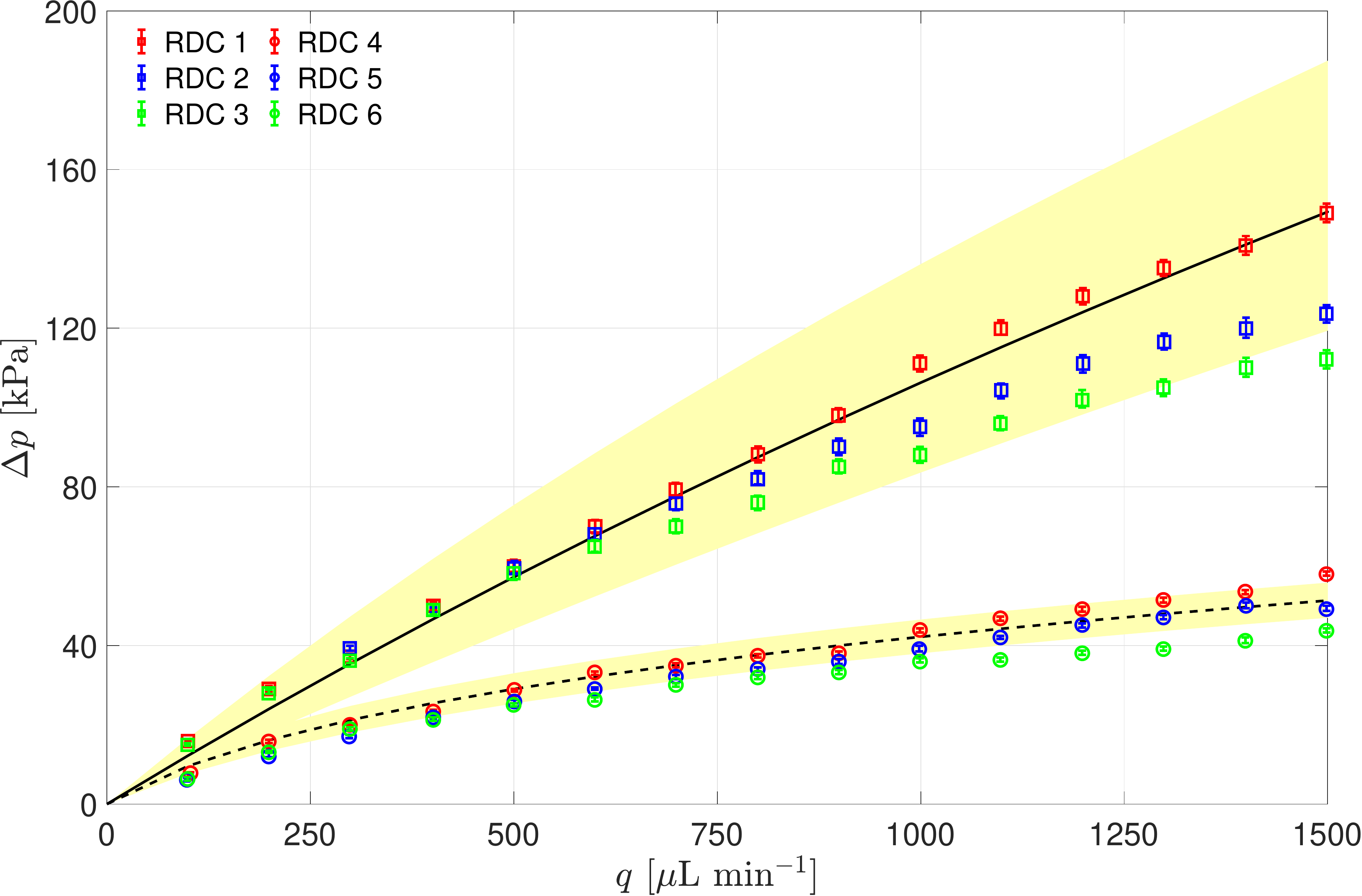}
    \caption{Comparison between our theory and the experimental data from \cite{RDC17} for the pressure drop $\Delta p$ as a function of the flow rate $q$. The symbols represent the experimental data for the different cases described in table~\ref{tab:RDC}. The curves are the predictions from equation \eqref{qdpthick}. The solid curve is for $E_Y=2.801$ MPa, while the dashed curve is for $E_Y=0.157$ MPa. The shaded region about each curve represents the uncertainty in $\Delta p$ due to the reported uncertainty in the undeformed channel height (i.e., $h_0 = 50 \pm 5\mu$m).}
    \label{fig:RDC_qdp}
\end{figure}

More recently, Raj \textit{et al.}\ studied the hydrodynamic resistance in microchannels by varying the top wall thickness as well as the Young's modulus \cite{RDC17}. Six sets of experimental data were reported, with parameters summarized in table \ref{tab:RDC}. A model based on the thick plate assumption was also proposed in \cite{RDC17}. Unfortunately, we have found that the model cannot explain the FSI because the top wall within the present thickness range cannot be regarded as the thick plate (recall section~\ref{sec:solids}). In figure \ref{fig:RDC_qdp}, we compare the flow rate--pressure drop relation from our theory to the experiments; once again, favorable agreement is observed. Note that equation \eqref{uyinf} does not involve the thickness $t$, which is why only one prediction curve is obtained for all the three thicknesses used in these experiments. The shaded region represents the 5 $\mu$m uncertainty in the underformed channel height reported in \cite{RDC17}. 

\begin{table}
    \centering
    \begin{tabular}{c c c c c c c}
            \hline
            Case & $t$ & $E_Y$ & $\delta$ & $\epsilon$ & $\gamma$ & $1/(\gamma\delta)^2$\\
              & [mm] & [MPa] & [\,--\,] & [\,--\,]  & [\,--\,]  &  [\,--\,]   \\
            \hline
            RDC 1 ({\color{red}$\square$}) & 2.0 & 2.801 & 0.1286--0.1571 & 0.0015--0.0018 & 36.36--44.44 & 0.0306 \\
            RDC 2 ({\color{blue}$\square$}) & 1.0 & 2.801 & 0.1286--0.1571 & 0.0015--0.0018 & 18.18--22.22 & 0.1225 \\
            RDC 3 ({\color{green}$\square$}) & 0.5 & 2.801 & 0.1286--0.1571 & 0.0015--0.0018 & 9.091--11.11 & 0.49~~~ \\
            RDC 4 ({\color{red}$\bigcirc$}) & 2.0 & 0.157 & 0.1286--0.1571 & 0.0015--0.0018 & 36.36--44.44 & 0.0306 \\
            RDC 5 ({\color{blue}$\bigcirc$}) & 1.0 & 0.157 & 0.1286--0.1571 & 0.0015--0.0018 & 18.18--22.22 & 0.1225 \\
            RDC 6 ({\color{green}$\bigcirc$}) & 0.5 & 0.157 & 0.1286--0.1571 & 0.0015--0.0018 & 9.091--11.11 & 0.49~~~ \\
            \hline
    \end{tabular}
    \caption{Values of the physical parameters used in the experiments of Raj \textit{et al.}\ \cite{RDC17}. The microchannel is $w=350$ $\mu$m wide, $l=3$ cm long and $h_0=50$ $\pm$ 5 $\mu$m in height. Based on the reported experimental conditions in \cite{RDC17}, the fluid viscosity is taken to be $\mu = 9.110 \times 10^{-4}$ Pa$\cdot$s, and the Poisson ratio is $\nu = 0.5$, for all the cases. }
    \label{tab:RDC}
\end{table}

It is well known that microfluidic measurements are highly sensitive to the channel height because $\Delta p \sim \mu l q/(wh_0^3)$ in the lubrication limit \cite{SSA04}. Almost all the experimental data fall into the shaded region, showing that the present theory is able to give quantitative prediction of the hydrodynamic resistance, but perhaps the experiments in \cite{RDC17} were not accurate enough to achieve their goal of addressing the effect of $t/w$. There are slight deviations in the cases RDC 3 and RDC 6 at large flow rates, which might be a consequence of the large elastic deformation in those cases. Also, these two cases have $1/(\gamma\delta)^2=0.49$, thus they are already the least favorable ones from the point of view of the limits of applicability of the proposed theory, which requires $1/(\gamma\delta)^2\ll1$. Overall, the agreement between the theoretical predictions and the experiments is quite satisfactory. 

\begin{figure}[t]
    \centering
    \includegraphics[width=0.75\textwidth]{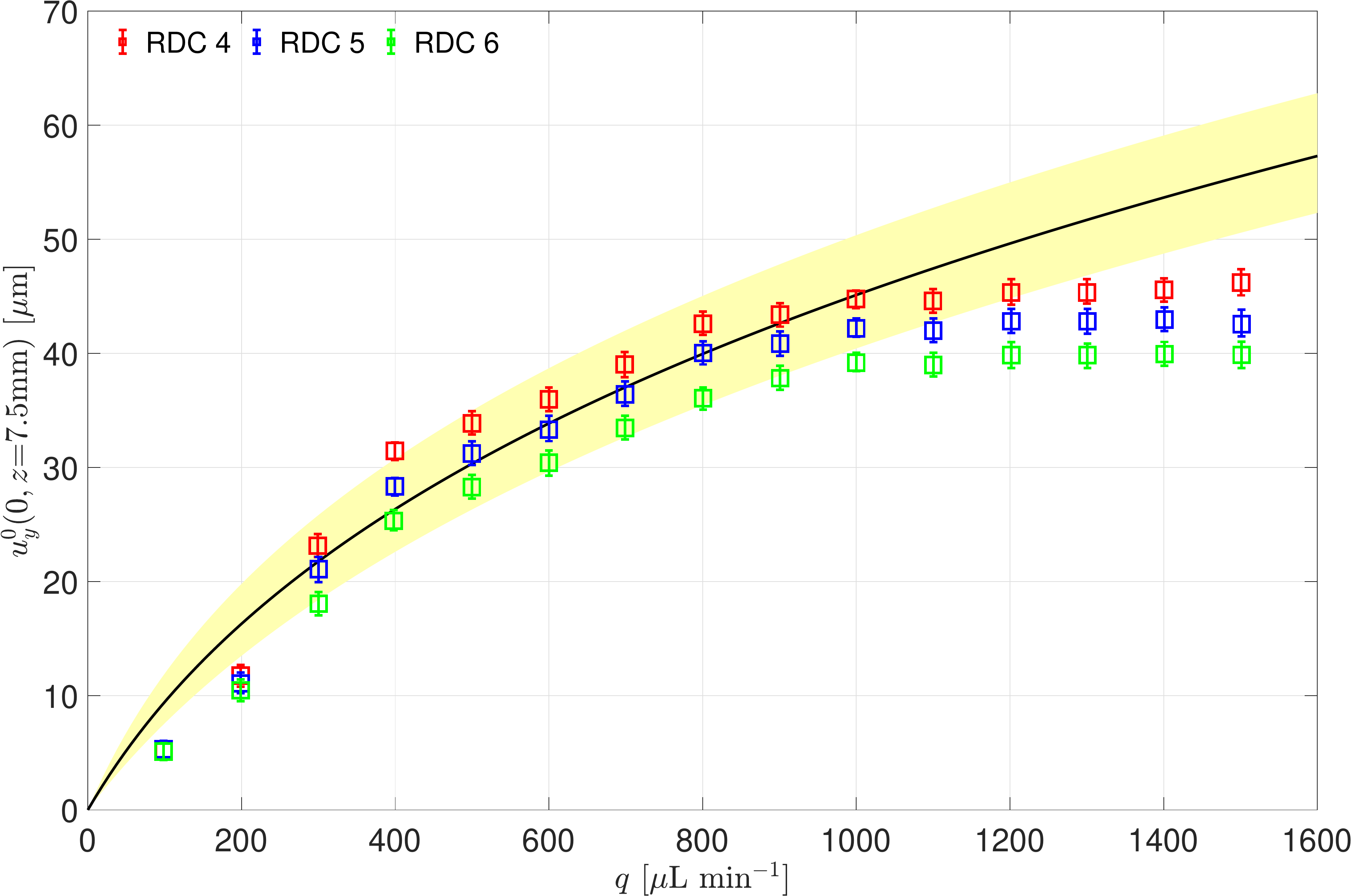}
    \caption{Comparisons of the maximum interface deflection $u_y^0(0,z)$ as a function of the flow rate $q$ at the $z=7.5 $ mm cross-section with $E_Y=0.157$ MPa between our theory and the experimental data from \cite{RDC17}. The symbols represents the reported experimental data, as described in table~\ref{tab:RDC}, while the solid curve is the prediction from equation \eqref{uyinf}. The shaded region represents the uncertainty in $\Delta p$ due to the reported uncertainty in undeformed channel height (i.e., $h_0 = 50 \pm 5$ $\mu$m).}
    \label{fig:RDC_uy}
\end{figure}

We also compare the maximum deflection of the fluid--solid interface at $z=7.5$ mm, with $E_Y=0.157$ MPa, as a function of the flow rate. Figure \ref{fig:RDC_uy} shows a comparison between the theoretical prediction and the experimental data from \cite{RDC17}. The agreement is best for the smaller flow rates. At larger flow rates, the experiment suggest that the deformation saturates, i.e., stops increasing, unlike the  theoretical prediction. In this case, we believe that nonlinear elastic effects, which our linear elastic theory cannot capture, begin to dominate. Another two sets of experimental data for $z=15$ mm and $z=22.5$ mm were also provided in \cite{RDC17}. However, we believe that there are potentially some misprints in \cite{RDC17} because according to the previous discussions, the maximum displacement at the fluid--solid interface is expected to be a concave, instead of a convex, function of $z$  (see figures \ref{fig:surf} and  \ref{fig:DPUY}). Furthermore, in spite of the deviations of the interface deflection at large $q$, we find that the prediction of the hydrodynamic resistance is still good (see figure \ref{fig:RDC_qdp}), which means that the flow rate--pressure drop relation is not sensitive to discrepancies in the maximum deformation, and so it can be pushed to a larger range of flow rates than one might \textit{a priori} expect.

\section{Conclusion}
\label{sec:conclusion}

In this work, we presented a theoretical study of the steady-state fluid--structure interaction between a Newtonian fluid and a long and shallow microchannel with a thick compliant top wall. This problem arises in a variety of lab-on-a-chip applications in which soft polymers, such as PDMS, are used to manufacture flow conduits. However, a complete mathematical theory of the ``bulging effect'' (flow-induced deformation) observed in experiments has remained elusive. To this end, under the lubrication approximation, we appealed to the standard result that the axial velocity profile is parabolic in any cross-section, even if the cross-section varies in the flow-wise direction. Then, importantly, we coupled the fluid and solid mechanics problems through the deformation of the fluid--solid interface (bulging of the microchannel's top wall). 

A scaling analysis of the elastostatics equations \eqref{solid_stress} for the solid showed that the stresses in the cross-sections perpendicular to the flow-wise direction are dominant and, thus, the 3D solid mechanics problem is simplified to a 2D plane strain problem. Assuming small strains and using the linear theory of elasticity, we show that the top wall's thickness plays an important role in the stress distribution in solid and, accordingly, has a significant influence on the boundary conditions to be imposed at two lateral surfaces of the top wall. By requiring that width $w$ and thickness $t$ are such that $1/(\gamma\delta)^2 = (w/t)^2 \ll 1$, also defined as the large-thickness case in the present study, the top wall deformation was decoupled in the flow-wise direction, allowing us to treat it as a simply supported rectangle at each cross-section. This analysis yielded a self-similar deflection curve at the fluid--solid interface, when scaling the deformation by the pressure. Furthermore, the present analysis showed that the characteristic scale for the interface deformation for the thick-wall problem is independent of thickness as in \cite{GEGJ06} (but different from the plate-like problem \cite{CCSS17,SC18,RCDC18}), specifically $\mathcal{P}_0 w/\overline{E}_Y = \mathcal{P}_0 w(1-\nu^2)/{E_Y}$, which is the expression from \cite{GEGJ06}, $\mathcal{P}_0 w/E_Y$, corrected for a plane strain configuration.

Integrating the flow velocity at a cross-section, we obtained flow rate--pressure drop relation, which deviates from the Poiseuille's law because it nonlinearly depends on the compliance of the top wall. The results predicted by the present theory agree favorably with the previous experimental studies \cite{GEGJ06, RDC17}. While previous theoretical analyses \cite{CCSS17,SC18,STGB17,BGB18} have successfully addressed this type of fluid--structure interaction for thinner plate-like top walls with $t/w\lesssim1$, the present theory is the first to quantify the hydrodynamic resistance in shallow compliant microchannels with \emph{thick} top walls such that $(t/w)^2\gg1$. 

The present theory is not only fitting-parameter-free but also directly solves for the fluid--solid interface deflection profile without assuming any specific shape. Our theory uncovers the physics hidden in the fitting parameter, $\alpha$, of the widely used model \eqref{gervais}, of which many \emph{ad-hoc} variations have been proposed \cite{CTS12,RS16,RDC17}. The present analysis also provides a clear answer for why the previous plate-theory-based models \cite{CCSS17,SC18,MY19} \emph{cannot} be pushed to large-thickness regime (even qualitatively) by showing that the bending effects are trivial in the present model. The differences between these theories are also reflected by the different parameter dependencies of the dimensionless numbers quantifying compliance.

Triggered by the observed deviations from the experiments, a remaining question is nonlinear effects, which arise at larger flow rates. {One open question is whether large deformations, which occur at higher flow rates, invalidate our assumption of plane strain, perhaps due to significant stretching along the flow-wise direction (see also the discussion of the ``RS'' data set in \cite{SC18}), which means that the cross-sectional deformation profiles are no longer decoupled in the flow-wise direction.} Additionally, PDMS is known to be porous, thus uptake solvents and swell \cite{LPW03}, which might necessitate considering large-deformation \emph{poroelastic} effects, as in \cite{PH13,AM17}.

Another interesting case is when the thickness of top wall of the microchannel is reduced until it can be considered to be ``thin.'' We provide a preliminary analysis in Appendix, and we argue that the present solid mechanics problem herein should be reduced to a simply supported beam with tension instead of the clamped beam, as in the previous studies using plate theories \cite{CCSS17,SC18}. Unfortunately, no experimental investigations on this configuration are available to validate the theory against. It would be of interest to conduct direct numerical simulations in future work to further delve into this problem.

\dataccess{The experimental data used is available in the references cited. All plots herein were generated from the equations in the text. No additional data was created in the course of this study.}
\aucontribute{Both authors contributed to all aspects of this work.}
\competing{We declare that we have no competing interests.}
\funding{This research was supported, in part, by the U.S.\ National Science Foundation under grant No.\ CBET-1705637 (to ICC) and a Ross Fellowship from The Graduate School at Purdue University (to XW).}

\bibliographystyle{rspublicnatwithsort.bst}
\setlength{\bibsep}{0pt}
\bibliography{references.bib,other.bib}

\clearpage
\setcounter{page}{1}
\appendix

\begin{center}
\color{jobcolor}
{\LARGE
\textsf{Supplementary Appendix for}\\
\textsf{``Theory of the flow-induced deformation of shallow compliant microchannels with thick walls''}}

\bigskip
\textsf{\Large by Xiaojia Wang and Ivan C.\ Christov}
\end{center}

\section{Moderate thicknesses and effect on the boundary conditions}
\label{appx}

Consider a configuration similar to that shown in figure \ref{fig:schematic} but the thickness of the top wall is not as large. (The side walls are still considered very thick, specifically infinite for the purposes of this discussion.) In this moderate-thickness case, the plane strain assumption is valid as long as the deformation is small enough, but the boundary conditions imposed at the sidewalls are not clear because $\sigma_{xx}$ is not negligible. Specifically, it is not necessarily correct to impose the simply supported boundary conditions from section~\ref{sec:solids}. At the same time, there is no good reason to impose clamped boundary conditions, as the previous studies on microchannels with thinner, plate-like walls \cite{OYE13, CCSS17, SC18}, because for the geometry considered herein the side surfaces are allowed to deform, while they were assumed to be rigid in those previous studies.

\begin{figure}[hb]
    \centering
    \includegraphics[width=0.8\textwidth]{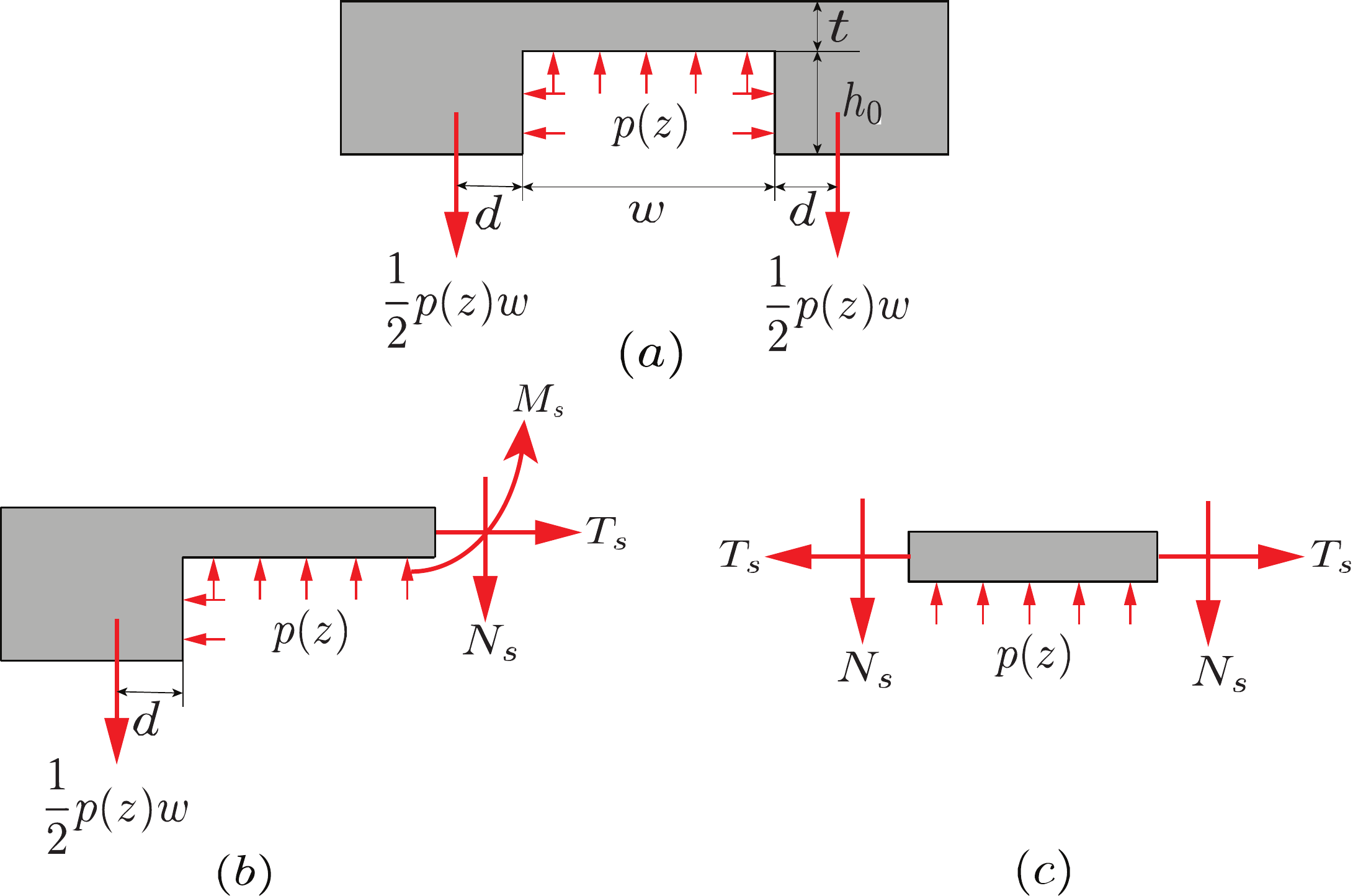}
    \caption{The force systems in the  cross-section of the elastic solid's wall for moderate thickness.}
    \label{fig:cross_sec}
\end{figure}

To understand the type of support at the side walls in the moderate-thickness case, we start from the free body diagram in figure \ref{fig:cross_sec}(b). Then, the reaction forces at the sidewalls are
\begin{equation}\label{side_forces}
    T_s = p(z)h_0,\qquad N_s = \frac{1}{2}p(z)w,\qquad M_s = \frac{1}{2}p(z)(h_0^2-wd),
\end{equation}
where $T_s$, $N_s$, and $M_s$ denote the tension, shear force and the moment respectively. Here $d$ is introduced to represent the point of the reaction force at the bottom of the side solid. Since $h_0\ll w$, we expect that $d\ll w$ due to stress concentration. Within the top wall, the resultant tension, $T$, shear force, $N$, and moment, $M$, are expected to be: $T\sim T_s$, $N\sim N_s$, $M \sim p(z)w^2/2+M_s \sim p(z)w^2(1+\delta^2-d/w)/2 \sim p(z)w^2/2$. Hence, we neglect $M_s$ in the following analysis. 

Thus, we can consider the configuration from figure \ref{fig:cross_sec}(c), a slender rectangle subject to pressure at the bottom, and shear and tension forces at the sidewalls. Note that the Airy stress function is still applicable in this case, but it is challenging to solve the corresponding biharmonic equation \eqref{airy} with the inclusion of tension. Fortunately, the thinness of the structure makes Saint-Venant's principle applicable, which states that ``a local force system has negligible effect on the stress distribution at distances that are large compared with the dimension of the surface where the forces are applied'' \cite{TG51}. Accordingly, the displacement field can be estimated based on a classic engineering model, without knowing the exact details of the stress distribution in solid. Thus, we regard the top wall as a simply supported beam with tension and solve for the displacement field by extending Timoshenko's beam theory.

Mechanical equilibrium requires that
\begin{equation}
    \frac{\partial T}{\partial x} =0,\qquad
    T\frac{\partial^2 u^0_y}{\partial x^2}+\frac{\partial N}{\partial x}+p(z) =0,\qquad
    -\frac{\partial M}{\partial x}+N =0
\label{eq:app_eq}\end{equation}
where $u^0_y$ now represents the deflection of the mid-plane of the beam ($y=t/2$). The deformation at the fluid--solid interface is believed to be very close to that of the mid-plane due to the slenderness of the top wall. The corresponding constitutive relations are
\begin{equation}
    M = -\overline{E}_YI\frac{\partial \varphi}{\partial x},\qquad
    N = \varkappa tG \left(-\varphi+\frac{\partial u_y^0}{\partial x}\right),
\label{eq:app_consti}
\end{equation}
where $I=t^3/12$ is the second area moment of the beam cross-section, $\varphi$ represents the rotation of the normal of the cross-section, and $\varkappa$ is the shear correction factor \cite{CE19}. As before, assuming zero displacement, as well as negligible moment at $x=\pm w/2$, the boundary conditions are
\begin{equation}\label{bcthin}
    u_y^0|_{x=\pm w/2} = 0,\qquad  M|_{x=\pm w/2} = 0.
\end{equation}
 
Equation \eqref{eq:app_eq}$_1$ shows that the tension is constant in the cross-section, i.e., $T=T_s=p(z)h_0$. Then, equations ~\eqref{eq:app_eq}~--~\eqref{eq:app_consti} can be rewritten in terms of $\varphi$ and made dimensionless:
\begin{equation}\label{phi_dim}
    \frac{\partial^4 \varphi}{\partial X^4}-\zeta P(Z)\left[\frac{\partial^2 \varphi}{\partial X^2}-\frac{(1+\overline{\nu})}{6\varkappa}\left(\frac{t}{w}\right)^2\frac{\partial^4 \varphi}{\partial X^4}\right]=0.
\end{equation}
Here the constant $\zeta=Tw^2/(\overline{E}_YI)=\mathcal{P}_0h_0w^2/(\overline{E}_YI)$ has been introduced to quantify the tension effect. In equation~\eqref{phi_dim}, the first term represents the bending effect. The terms in the bracket represents the influence of tension, and the thickness effect is captured by the second term. Given the typical range of parameters for a microchannel, we conclude that the tension cannot be neglected here. Therefore, for the small thickness case, the top wall can no longer be regarded as a simply supported rectangle but, rather, it behaves like a beam with an immovable edge, i.e., simple support plus tension \cite{TWK59}.

Soving equations ~\eqref{eq:app_eq}~--~\eqref{eq:app_consti}, the vertical displacement of the mid-plane, which is also approximately the vertical displacement at the fluid--solid interface, is found to be
\begin{equation}\label{uybeam}
    u_y^0(x,z) = \left[\frac{w^2}{4\mathfrak{u}^2(z)h_0}-\frac{(1+\overline{\nu})t^2}{6\varkappa h_0}\right]\left\{\frac{\cosh{[2\mathfrak{u}(z)x/w]}}{\cosh{\mathfrak{u}(z)}}-1\right\} - \frac{1}{2h_0}\left(x+\frac{w}{2}\right)\left(x-\frac{w}{2}\right),
\end{equation}
where 
\begin{equation}
    \mathfrak{u}^2(z) = \frac{p(z)h_0 w^2}{4\overline{E}_Y I}\left[\frac{1}{p(z)h_0/(\varkappa t G)+1}\right] = \frac{\zeta P(Z)}{4}\left[1 + \frac{(1+\overline{\nu})}{6\varkappa}\left(\frac{t}{w}\right)^2\zeta P(Z)\right]^{-1}.
\end{equation}
If $(t/w)^2\ll 1$, then $\mathfrak{u}^2(z)\approx\zeta P(Z)/4 = p(z)h_0 w^2/(4\overline{E}_Y I)$ and the second term in the bracket in equation \eqref{uybeam} also vanishes, then equation \eqref{uybeam} is reduced to the Euler--Bernoulli beam with tension \cite{TWK59}. However, unlike equations \eqref{disp}$_2$ and \eqref{uyinf} of the large-thickness case, the deflection profile in equation \eqref{uybeam} no longer displays self-similarity along the flow-wise direction because $p(z)$ cannot be factored out.

\begin{figure}
    \centering
    \includegraphics[width=0.75\textwidth]{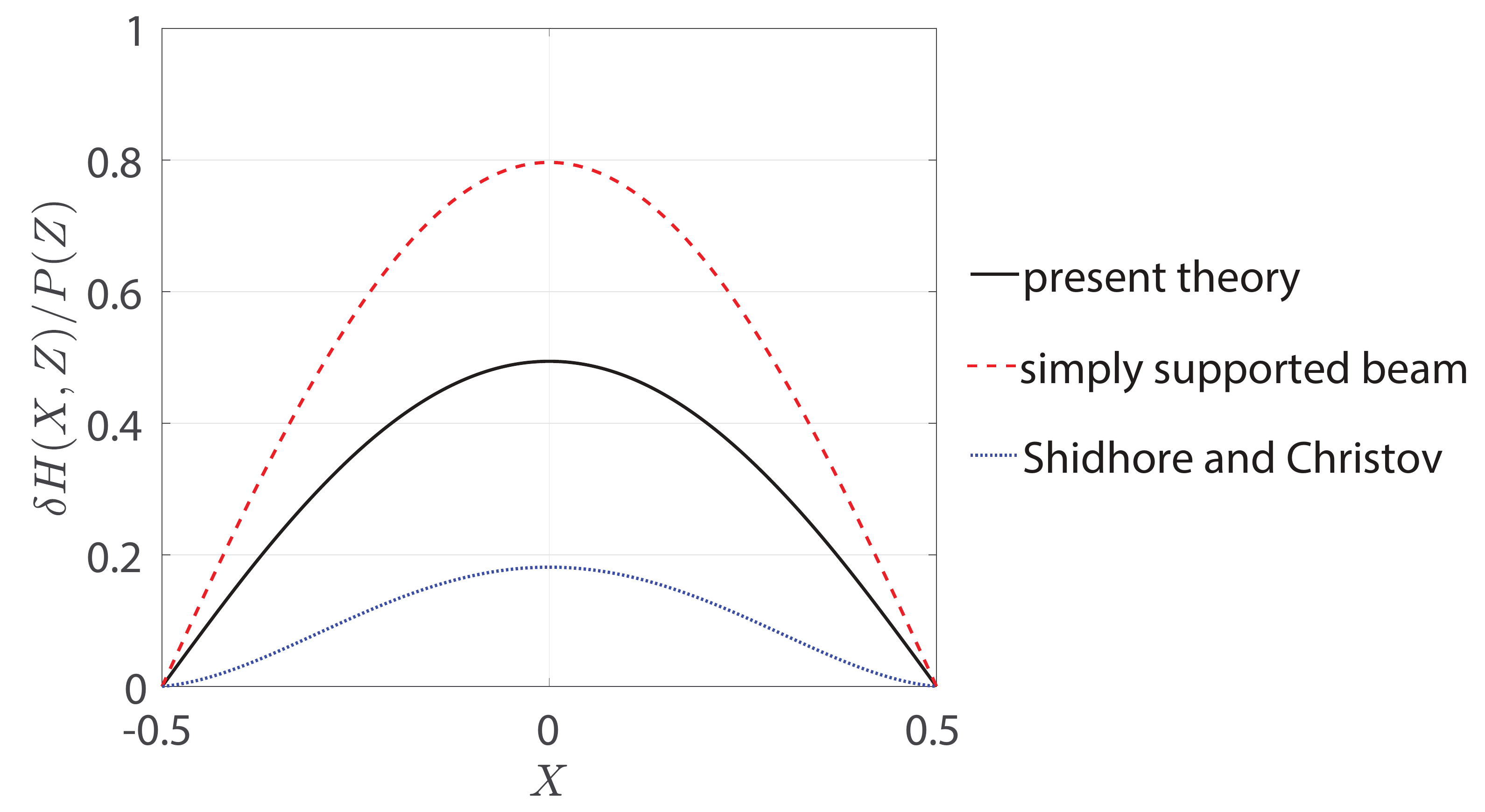}
    \caption{The cross-sectional deformation profile $\delta H(X,Z)/P(Z)$ versus $X$ for a microchannel with $\delta =0.1$ and $t/w = 0.1$ under a pressure drop such that $\zeta = 5$. The solid curve corresponds to equation \eqref{uybeam}. The dashed curve represents the deformation of a simply supported beam using equation \eqref{disp}$_2$. For comparison, the dotted curve is the solution from Shidhore and Christov \cite{SC18} for a clamped thick-plate-like wall.}
    \label{fig:disp_thin}
\end{figure}

To get a sense of the tension effect, we compare this proposed moderate-thickness theory with other models in figure \ref{fig:disp_thin}. Note that we compare $\delta H(X,Z)/P(Z)$, with $\delta H(X,Z) = u_y^0(x,z)/w$, to show the magnitude of the deformation. For the applicability of  linear elasticity, we expect that $\delta H(X,Z)\ll 1$. We can see that  tension suppresses deformation,  compared to the case of simply supported beam. However, the tension is not as restrictive as the clamping considered in  \cite{SC18}. 

As before, the flow rate--pressure drop relation is obtained by integrating the axial fluid velocity across the cross-section. Then, we rewrite $p(z)$ as a function of $\mathfrak{u}(z)$, so that equation \eqref{qpdiff} can be written entirely in terms of $\mathfrak{u}$:
\begin{equation}
    p(z)= \left[ \frac{h_0w^2}{4\overline{E}_Y I\mathfrak{u}^2(z)}-\frac{1}{\varkappa\gamma G} \right]^{-1} \quad\Rightarrow\quad q=-\frac{1}{12\mu}\frac{\mathrm{d}\mathfrak{u}}{\mathrm{d}z}
    \underbrace{\frac{\mathrm{d}p}{\mathrm{d}\mathfrak{u}}\int_{-w/2}^{+w/2}\left[h_0+u_y^0(x,z)\right]^3 \,\mathrm{d}x}_{=\mathcal{R}(\mathfrak{u})}.
\end{equation}
Using separation of variables, the solution of the last ODE can be expressed as a quadrature:
\begin{equation}\label{qdpthin}
    q = \frac{1}{12\mu(l-z)}\int_0^{\mathfrak{u}(z)} \mathcal{R}(\mathfrak{u'})\,\mathrm{d}\mathfrak{u'},
\end{equation}
where $\mathfrak{u}'$ is a ``dummy'' integration variable. Unfortunately, this integral can only be evaluated numerically. 

\end{document}